\documentclass[a4paper, 12pt]{article}

\usepackage{amsmath, enumerate, amsthm, amssymb, mathrsfs, graphicx, multirow, listings, lipsum, color, hyperref, cite, natbib, adjustbox, algpseudocode, ulem, multicol, lscape, parskip}
\usepackage[table]{xcolor}

\newcommand{\disp}{\displaystyle}

\begin{document}

\title{\textbf{Correlation-Adjusted Simultaneous Testing for Ultra High-dimensional Grouped Data}}
\author{
Iris Ivy Gauran$^{1}$, 
Patrick Wincy Reyes$^{2}$, 
Erniel Barrios$^{3}$,
Hernando Ombao$^{1}$}
\vspace{-3em}
\date{\small 
$^{1}$ Statistics Program, King Abdullah University of Science and Technology (KAUST)\\ 
$^{2}$ School of Statistics, University of the Philippines, Diliman\\
$^{3}$ School of Business, Monash University Malaysia}
\maketitle

\begin{abstract}
Epigenetics plays a crucial role in understanding the underlying molecular processes of several types of cancer as well as the determination of innovative therapeutic tools. To investigate the complex interplay between genetics and environment, we develop a novel procedure to identify differentially methylated probes (DMPs) among cases and controls. Statistically, this translates to an ultra high-dimensional testing problem with sparse signals and an inherent grouping structure. When the total number of variables being tested is massive and typically exhibits some degree of dependence, existing group-wise multiple comparisons adjustment methods lead to inflated false discoveries.  We propose a class of Correlation-Adjusted Simultaneous Testing (CAST) procedures incorporating the general dependence among probes within and between genes to control the false discovery rate (FDR). Simulations demonstrate that CASTs have superior empirical power while maintaining the FDR compared to the benchmark group-wise. Moreover, while the benchmark fails to control FDR for small-sized grouped correlated data, CAST exhibits robustness in controlling FDR across varying group sizes.  In bladder cancer data, the proposed CAST method confirms some existing differentially methylated probes implicated with the disease (Langevin, et. al., 2014). However, CAST was able to detect novel DMPs that the previous study (Langevin, et. al., 2014) failed to identify. The CAST method can accurately identify significant potential biomarkers and facilitates informed decision-making aligned with precision medicine in the context of complex data analysis.
\end{abstract}
\vspace{1em}
\small{
\textbf{Keywords:}
False Discovery Rate, Group-wise Multiple Testing, Ultra High-dimensional data, Epigenome-wide Association Studies, Bladder Cancer
\vspace{1em}

\section{Epigenome-wide Association Studies on Cancer}
\label{sec:Introduction}

While genome-wide association studies (GWAS) have significantly advanced our understanding of genetic susceptibilities to various complex diseases, the DNA sequence variations identified through GWAS are estimated to account for less than 30\% of phenotypic variability in humans \citep{lander2011initial}. In fact, these variations may explain only a small portion of familial cancer risk \citep{stadler2010genome}. This unexplained variability has prompted increasing attention to epigenetic mechanisms, serving as a catalyst for the development of epigenome-wide association studies (EWAS) \citep{rakyan2011epigenome, esteller2008epigenetics, sharma2010epigenetics}. Epigenetic modifications, which are essential for the regulation of gene expression, influence fundamental cellular processes such as proliferation, differentiation, and apoptosis, and their dysregulation may lead to significant diseases, including cancer \citep{liep2012feedback, baylin2016epigenetic, dawson2012cancer}.  In EWAS, atypical and aberrant DNA methylation patterns can potentially serve as valuable cancer biomarkers, offering promising tools for early detection, diagnosis, prognosis assessment, and predicting therapy responses \citep{esteller2008epigenetics, laird2005cancer,ronn2015dna, bergman2013dna, yang2010targeting, jeronimo2014epigenetic}. As such, ongoing epigenetic and epigenomic research is essential to understanding how cancer and its complications develop through the combined effects of DNA sequence variations and environmental exposures that alter cellular phenotypes \citep{rosen2018epigenetics}.  Driven by the scientific investigation on the association between epigenetic mechanisms and cancer, the overarching goal of this study is to develop novel correlation-adjusted simultaneous testing (CAST) procedures for ultra high-dimensional grouped data to identify differentially methylated probes (DMPs) while controlling the proportion of false positives.

Bladder cancer (BC) provides a compelling case for exploring the role of epigenetic modifications in cancer development and progression \citep{enokida2008epigenetics,langevin2014leukocyte,martinez2019epigenetics}. Several cancer-related genes have been linked to the progression and prognosis of BC through gene-expression profiling, revealing differential regulation in BC compared to normal tissue \citep{mitra2006molecular, kawakami2006identification, kawakami2007increased}. Understanding the molecular mechanisms driving BC progression, including genetic alterations and epigenetic changes, is critical as these heritable modifications in gene expression occur without changes to the underlying DNA sequence \citep{wolffe1999epigenetics}.  Among the various epigenetic modifications, DNA methylation has been the most extensively studied in the context of bladder cancer \citep{porten2018epigenetic}. Clinically, a viable tumor biomarker must be detectable through minimally invasive procedures \citep{ransohoff2003developing}. DNA methylation fulfills this criterion and shows promise as a biomarker for several reasons \citep{ransohoff2003developing,herman2003gene}. Unlike genetic mutations, methylation occurs in defined regions, such as CpG islands, and can be detected with high sensitivity using techniques like methylation-specific PCR (MSP) \citep{herman1996methylation} and high-resolution bisulfite sequencing \citep{susan1994high}. Furthermore, hypermethylated DNA is associated with numerous types of tumor \citep{costello2000aberrant}, each having its unique signature of methylated genes. For instance, the GSTP1 gene is frequently methylated in prostate cancer \citep{santourlidis1999high,lee1997cg}, the VHL gene in renal cancer \citep{herman1994silencing,battagli2003promoter}, the MLH1 gene in colon cancer \citep{ricciardiello2003frequent}, and the APC gene in esophageal cancer \citep{kawakami2000hypermethylated}. However, no specific gene promoter hypermethylation uniquely identifies BC at this time \citep{enokida2008epigenetics}.  Therefore, this study addresses the need by developing a procedure for identifying DMPs among bladder cancer cases and healthy controls using an appropriate multiple comparison adjustment. Our proposed CAST methods demonstrate that DNA methylation can serve as a biomarker for bladder cancer risk, identifying novel gene targets that may play roles in disease progression and therapy. The CAST methods were able to streamline the analysis process, reducing the need to examine tens of thousands of CpG loci and genes, thereby saving significant time, effort, and resources in future research.

Among EWAS studies, the data collected include a small number of subjects relative to the information available per subject, such as millions of genetic variants or methylation profiles from approximately 850,000 probes \citep{pidsley2016critical}.  Statistically, this highlights that epigenetic data are high-dimensional or even ultra high-dimensional in nature.  With the deluge of available epigenetic data, identifying relevant features or differentially methylated probes poses a risk of committing inflated false discoveries. Many multiple testing procedures, like the Bonferroni correction and \cite{benjamini1995controlling} method, assume independence of test statistics and $P$-values, which can be problematic when dealing with ultra high-dimensional data. The method developed by \cite{benjamini2001control}, while addressing positive dependence, struggles with negative associations.  This study overcomes the serious limitations of existing methods by providing a unified framework Correlation-Adjusted Simultaneous Testing (CAST) which takes into account the inherent dependence structure from the data, within and between groups. The key advantages of our proposed method are: (1.) It incorporates both positive and negative correlations and captures the imbalance among the group sizes. (2.) CAST methods incorporate thresholds that result in both linear and quadratic step-up procedures. This study provides an adaptive and non-linear step-up procedure for a general dependence structure. (3.) CAST methods are robust, flexible and can be implemented to any collection of unadjusted $P$-values corresponding to null hypotheses tested simultaneously.  On the clinical aspect, identifying reliable biomarkers in bladder cancer could significantly reduce the need for invasive tests, enhance early detection, facilitate monitoring of treatment response, and provide critical information for determining prognosis and tailoring treatment intensity, thereby greatly benefiting patients and clinicians.

The remainder of this paper is organized as follows. In Section \ref{sec:method}, we discuss our proposed class of Correlation-Adjusted Simultaneous Testing (CAST) procedures which incorporates a data-adaptive $P$-value adjustment within and between groups. Important features of CAST method over the benchmark methods are visually characterized. In Section \ref{sec:results}, we present the results of our empirical comparison between CAST and the existing group-wise methods for both the simulated and bladder cancer data \citep{langevin2014leukocyte}. We summarize our contributions including our key findings, limitations, and the future directions of this paper in Section \ref{sec:conclusion}. 
\vspace{1.5em}

\section{Group-wise Multiple Testing Procedures controlling False Discovery Rate}
\label{sec:method}

\cite{benjamini2000adaptive} introduced an adaptive step-wise procedure that builds upon their foundational method \citep{benjamini1995controlling} to control the false discovery rate (FDR) for independent test statistics. \cite{benjamini1995controlling} demonstrated that the bound of their linear step-up multiple testing procedure is sharp when all null hypotheses are true, provided the test statistics are independent and continuous.   However, when some null hypotheses are false, the procedure becomes conservative, with the degree of conservatism proportional to $M_0/M$, where $M_0$ represents the number of true null hypotheses out of the total $M$ hypotheses \citep{benjamini2006adaptive}.  Notably, in any practical data analysis, the exact number of true null hypotheses, $M_0$, is typically unknown and must be estimated.

In the context of ultra high-dimensional data, where the total number of variables $M$ being tested is massive and typically exhibits some degree of dependence, the multiple comparisons adjustment methods proposed in \cite{benjamini1995controlling, benjamini2000adaptive} can be overly conservative, leading to a loss of statistical power.  To overcome this serious limitation, a more effective approach involves leveraging the inherent grouping of variables and performing the adjustment among the $M_g$ pairs of hypotheses within each group $g$, where $g \in [G] = {1, 2, \ldots, G}$. This strategy helps to account for the dependence among variables both within and between groups.
Using the Lowest SLope (LSL) method to estimate the proportion of true nulls, $\widehat{\pi}_{0g} = M_{0g}/M_g$, the grouped \cite{benjamini2000adaptive} (GBH) procedure will reject the null if the $j$th ordered $P$-value in the $g$th group satisfies
\begin{equation}
    \label{eqn:GBH}
    P_{g(j)} \leq \alpha_{\text{GBH}} \equiv \disp \frac{\alpha j}{ \widehat{\pi}_{0g} M_g}.
\end{equation}
However, a key limitation of several multiple testing procedures including the Bonferroni correction, \cite{benjamini2000adaptive} and \cite{benjamini1995controlling} method is that they rely on the independence assumption.  Thus, for correlated test statistics and $P$-values, \cite{benjamini2001control} showed that false discoveries can also be controlled under positive regression dependence structure.  Using the grouped \cite{benjamini2001control} (GBY) procedure, the method will reject the null hypothesis if
\begin{equation}
    \label{eqn:GBY}
    P_{g(j)} \leq \alpha_{\text{GBY}} \equiv \disp \frac{\alpha j}{ \widehat{\pi}_{0g} M_g \mathcal{C}(M_g)} = \frac{\alpha_{\text{GBH}}}{\mathcal{C}(M_g)},
\end{equation}
where $\mathcal{C}(M_g) = \sum \limits_{j=1}^{M_g} \frac {1}{j}$ is the harmonic mean. 

However, in EWAS studies, probes belonging in the same gene can display negative correlations \citep{gomez2021dna}. As noted, this can be seriously problematic in other studies which also display a negative associations among the variables being tested because \eqref{eqn:GBY} is restricted only to a positive dependence structure \citep{benjamini2001control}.  This study addresses the serious limitation by adaptively characterizing a general dependence structure and incorporating it in the adjustment factor. We demonstrate in Section \ref{sec:results} that the adjustment factor is able to control the rate of false positives with desirable empirical power.  
\vspace{1.5em}

\subsection{Correlation-Adjusted Simultaneous Testing (CAST)}
\label{sec:CAST}

In this section, we develop a novel method that conducts the multiple testing adjustment in two stages. Within the $g$th group, let $\mathbf{r}_j = (r_{j1}, \ldots, r_{jj}, \ldots, r_{jMg})^\top$ be the vector of correlations between the $j$th and $j'$th variable, $j' \in [M_g] = \{1, 2, \ldots, M_g\}$, $|r_{jj'}| \leq 1, ~r_{jj'} = 1 ~\text{when}~j = j'$. First, we perform the within group adjustment as follows
\begin{equation}
    \label{eqn:CASTWithin}
    \text{A}_{\text{W}} = \disp \frac{j}{\widehat{\pi}_{0g} M_g \mathcal{C}(M_g, \mathbf{r}_j) }.
\end{equation}
Second, we incorporate the between-group adjustment to \eqref{eqn:CASTWithin} as
\begin{eqnarray}
    \label{eqn:CASTBetween}
    \text{A}_{\text{B}} 
    &=& \frac{M_g}{\min\left(G M_g, M\right)} = \max \left(\frac{1}{G}, \frac{M_g}{M}\right) =
    \begin{cases}
        \disp \frac{1}{G} & \text{if}~ M_g < \disp \frac{M}{G}\\
        \disp \frac{M_g}{M} & \text{if}~ M_g \geq \disp \frac{M}{G}.
    \end{cases} \nonumber
\end{eqnarray}
The purpose of the adjustment in \eqref{eqn:CASTBetween} is to account for the varying group sizes $M_g, ~g \in [G]$.  Using the CAST method, we reject the null hypothesis if
\begin{equation}
    \label{eqn:CAST}
    P_{g(j)} \leq \alpha_{\text{CAST}} \equiv \alpha \text{A}_{\text{W}}\text{A}_{\text{B}} = \disp \frac{\alpha j}{\widehat{\pi}_{0g} \min\left(G M_g, M\right) \mathcal{C}(M_g, \mathbf{r}_j) }.
\end{equation}
The adjustment within and between groups in \eqref{eqn:CAST} implies that for a group with a large number of variables, e.g., a gene with numerous methylation probes, it is desirable for the adjustment to be more conservative to reduce the false discoveries. In contrast, the CAST method provides a boost for groups with lower number of variables to encourage discoveries.  This is suitable for the ultra high-dimensional case wherein the group information is integrated to the overall number of hypotheses to be tested.

Overall, CAST is a novel class of procedures, characterized distinctly by the adjustment factor $\mathcal{C}(M_g, \mathbf{r}_j)$ where the stages can be performed either sequentially as shown in \eqref{eqn:CASTWithin} and \eqref{eqn:CASTBetween} or concurrently as presented in \eqref{eqn:CAST}. 
Further, we propose a set of techniques where $f(\mathbf{r}_j) = \overline{r}_j$.  For the Linear CAST (LCAST) method, the correlation-adjusted factor is written as
\begin{eqnarray}
    \label{eqn:LCAST_C}
    \mathcal{C}(M_g, \overline{r}_j) &=& \sum\limits_{j = 1}^{M_g} \left(1 - f(\overline{r}_j) \right), \hspace{1em} \text{where}~~
    f(\overline{r}_j) = \frac{j - 1}{j + \overline{r}_j}, ~
    \overline{r}_j = \frac{1}{M_g} \sum \limits_{j' = 1}^{M_g} r_{jj'}.
\end{eqnarray}
Meanwhile, the correlation-adjusted factor for the Quadratic CAST (QCAST) method is 
\begin{equation}
    \label{eqn:QCAST_C}
    \mathcal{C}(M_g, \overline{r}_j) = M_g \left(1 - f(\overline{r}_j) \right) = M_g \left(\frac{1 + \overline{r}_j}{j + \overline{r}_j}\right).
\end{equation}
If $j = 1$ and $\overline{r}_j = -1$, we use $\mathcal{C}(M_g)$ in \eqref{eqn:GBY}. 
However, we observe that $\overline{r}_j = -1$ is highly unlikely in real data applications.  For instance, based on the nature of epigenetic data, there are sufficient probes to be unmethylated in the promoter region while there is a reversal in pattern in other parts of the gene \citep{anastasiadi2018consistent}.  Consider two specific probes $j$ and $j'$ such that they exhibit a strong inverse relationship, i.e. $r_{jj'}= -1$. However, the rest of the probes $j^\star$ in the same gene where the $j$th and $j'$th probe belongs to are not expected to lean on one extreme only. Thus, we do not expect the average correlation in the $j$th variable to be exactly $-1$.  Consequently, the proposed multiple comparisons adjustment adapts to the correlations of the group since the correlation values vary from one group to another.
\vspace{1em}

\noindent\textbf{Remarks:}
\begin{enumerate}
    \item 
    Consider the case when the total number of groups is 1. In the ungrouped setting, $M_g$ reduces to $M$, $g \in \{1\}$.  This indicates that \eqref{eqn:CAST} reduces to $P_{(j)} \leq \alpha \text{A}_{\text{W}} = \disp \frac{\alpha j}{\widehat{\pi}_{0} M \mathcal{C}(M, \mathbf{r}_j) }$.  In addition, if the correlation among variables is not utilized and we take $\mathcal{C}(M, \mathbf{r}_j = 0) = \sum\limits_{j = 1}^M \frac{1}{j}$, then the proposed CAST method is equivalent to the ungrouped adjustment of BY in \eqref{eqn:GBY}.
    
    \item 
    Consider the case when the group size is 1 for some groups. If some of the groups $g \in [G]$ contain only one variable, the estimate of $\widehat{\pi}_{0g}$ is set to 1 because the LSL method in \cite{benjamini2000adaptive} requires a minimum of two $P$-values.  Moreover, if $M_g = 1$ then $\mathbf{r}_j = (r_{jj}) = 1$ and $\mathcal{C}(M_g, \mathbf{r}_j) = 1$ .  This means that the adjustment presented in \eqref{eqn:GBH} and \eqref{eqn:GBY} both reduce to the classical ungrouped decision rule: we reject the null hypothesis if the $P$-value $\leq \alpha$. Conversely, our proposed method in \eqref{eqn:CAST} rejects the null hypothesis if the $P$-value $\leq \frac{\alpha}{G}$. 

    \item 
    Consider the case when the group size is 1 for all groups.  If each group satisfies 
    $M_g = 1 ~\forall g \in [G]$ then $G = M$.  The adjustment shown in \eqref{eqn:GBH} and \eqref{eqn:GBY} still rejects the null hypothesis if the $P$-value $\leq \alpha$ whereas the proposed CAST method
    converges to the Bonferroni correction which rejects the null hypothesis if the $P$-value $\leq \frac{\alpha}{M}$. 
\end{enumerate}
\vspace{-0.5em}

In Remark (1) the proposed and the existing BY method coincide when $G = 1$ and we don't take into consideration the correlations across variables.  From Remarks (2) and (3), GBH and GBY methods reduce to the classical testing without multiplicity correction. This indicates that existing group-wise methods produce a higher proportion of false positives for groups with only one variable. In general, this phenomenon of inflated FDR may be observed when the group sizes are small or the number of tests per group is small.

In this section, we reiterate that this study rests on the inherent grouping that the variables possess. When no such biological or scientific prior information is available, unsupervised grouping techniques such as clustering may be conducted as a pre-processing step. Note however, that this study can also be used in the ungrouped setting, i.e., $G = 1$ and $M_g = M_1 = M$.

Therefore, the primary \underline{advantage} of CAST methods is that they are robust and can be implemented to any collection of $M$ unadjusted $P$-values extracted using a $P$-value generating algorithm $\mathcal{A}$ associated to $\mathcal{H}_{0j}: \theta_j \in \Theta_0$, where $\Theta_0$ is the null space, $j \in [M_g], g \in [G], M_1 + M_2 + \ldots + M_G = M$.  Therefore, the CAST method addresses the problem of testing $M$ null hypotheses simultaneously, $\mathcal{H}_{0j}$ is true for $j \in [M]$.
\vspace{1.5em}

\subsection{Features of the Linear and Quadratic CAST Methods}
\label{sec:LCASTandQCAST}

To further elucidate the difference between CAST procedures and existing GBH and GBY procedures refer to Figure \ref{fig:CASTthreshold}. In particular, we demonstrate how the proposed thresholds perform in a small-sized group scenario as well as with varying magnitudes and directions of the average correlations. For the LCAST method with average pairwise correlation between all variables assumed to be constant within the group, i.e. $\overline{r}_j = \overline{r}$, for $1 < M_g \leq M < \infty$, we have
\begin{eqnarray}
\label{eqn:LimitCorrs}
    \lim \limits_{\overline{r} \rightarrow -1^+} \mathcal{C}(M_g, \overline{r}_j) 
    &=& \lim \limits_{\overline{r} \rightarrow -1^+} \sum \limits_{j = 1}^{M_g} \left(1 - \frac{j - 1}{j + \overline{r}}\right) \rightarrow \varepsilon > 0 \nonumber\\
    \lim \limits_{\overline{r} \rightarrow 0} \mathcal{C}(M_g, \overline{r}_j) &=& \sum \limits_{j = 1}^{M_g} \frac{1}{j} 
    \approx \ln(M_g) + \frac{1}{2M_g} + \gamma \nonumber\\
    \lim \limits_{\overline{r} \rightarrow 1^-} \mathcal{C}(M_g, \overline{r}_j) 
    &=& 1 + 2\sum \limits_{k = 1}^{M_g - 1} \frac{1}{k} \approx 2\ln(M_g - 1) + \frac{M_g}{M_g - 1} + 2\gamma \nonumber
\end{eqnarray}
where $\gamma$ is the Euler–Mascheroni constant.  Because the correlation-adjustment $\mathcal{C}(M_g, \overline{r}_j)$ is included in the denominator of $\alpha_{\text{CAST}} = \alpha \text{A}_{\text{W}}\text{A}_{\text{B}}$, then the results in \eqref{eqn:LimitCorrs} indicate that groups with positively correlated variables will have more stringent rejection thresholds.

\begin{figure}[!ht]
    \centering 
    \caption{The comparison of the rejection thresholds using the proposed CAST methods with several values of $\overline{r}_j = \overline{r}$ versus the grouped version of BH and BY procedures, for a fixed group. LCAST is utilized with (a) zero and negative values of $\overline{r}$ compared with (b) zero and positive values of $\overline{r}$. Similarly, QCAST is employed with (c) zero and negative values of $\overline{r}$ versus (d) zero and positive values of $\overline{r}$.}
    \label{fig:CASTthreshold}
    \begin{tabular}{cc}
    \includegraphics[width=0.45\linewidth]{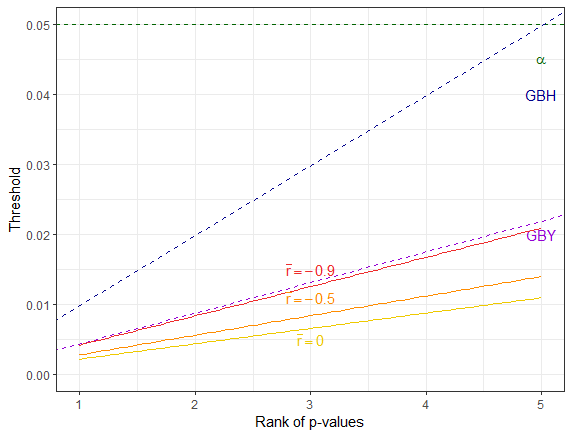} &
    \includegraphics[width=0.45\linewidth]{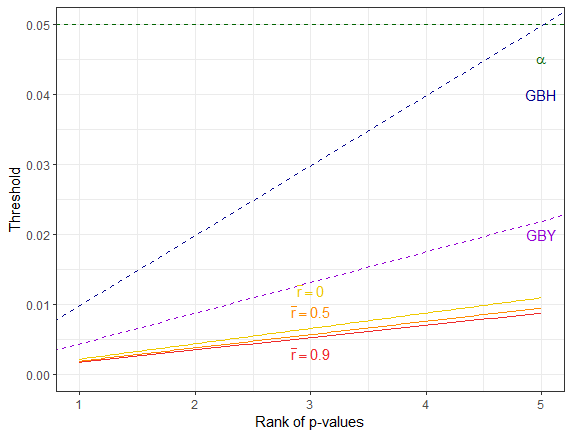} \\
    (a)  & (b) \\
    \includegraphics[width=0.45\linewidth]{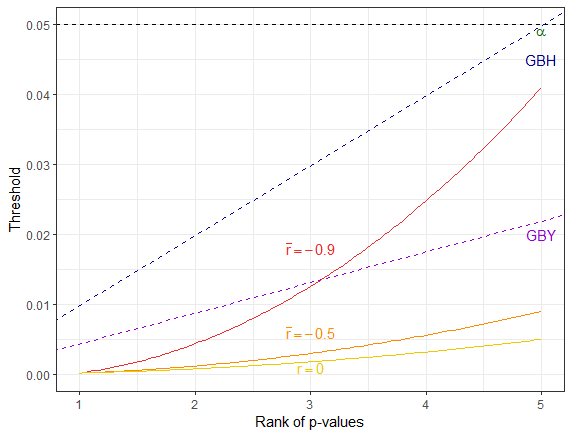} &
    \includegraphics[width=0.45\linewidth]{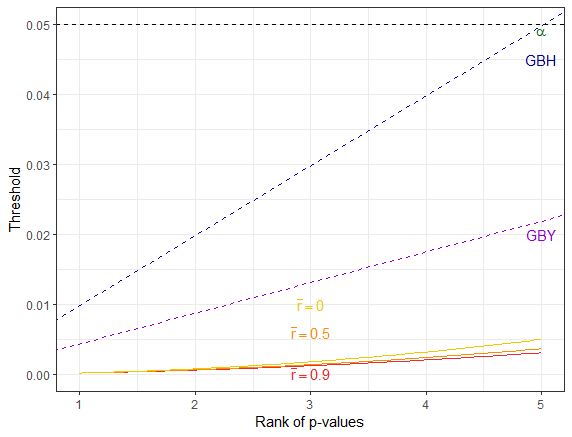} \\
    (c) & (d)\\
    \\
    \end{tabular}
\end{figure}

Consistent with the results in \eqref{eqn:LimitCorrs}, Figure  \ref{fig:CASTthreshold} shows that the rejection threshold varies according to the group mean correlation. Since the Benjamini-Hochberg and Benjamini-Yekutieli methods are linear step-up procedures, they both employ a linear threshold to determine the value of $r$, which indicates the rejection of $P_{(1)}, P_{(2)}, \ldots, P_{(r)}$, where $P_{(1)}, P_{(2)}, \ldots, P_{(M)}$ represent the ordered observed $P$-values \citep{benjamini1995controlling, benjamini2000adaptive, benjamini2001control, benjamini2006adaptive}.  This can be visualized in Figure \ref{fig:CASTthreshold} where the x-axis represents the indices $j$ from 1 to $M$, and the y-axis shows the ordered $P$-values. Starting with the largest $P$-value, $P_{(M)}$, we check whether it falls above or below the linear threshold. If it is below, we reject all hypotheses and set $r = M$. If the $P$-value falls below the threshold, reject all null hypothesis and set $r = M$. If the $P$-value is above the threshold then we move to $P_{(M - 1)}$ and repeat the process, continuing until we find the first $P$-value $P_{(r)}$ that is below the line, where $r$ is the corresponding index. As mentioned above, after determining $r$, the linear step-up procedure leads to the rejection of $P_{(1)}, P_{(2)}, \ldots, P_{(r)}$. Hence, when comparing linear thresholds, a higher slope indicates a more liberal test.

From Figure \ref{fig:CASTthreshold}, the slope of the rejection threshold for LCAST and QCAST is highest at strong negative correlations and lowest at strong positive correlations. This means that CAST methods will yield more rejections among negatively correlated variables. We demonstrate in the simulation experiments in Section \ref{sec:NumericalStudies} that these are correct rejections and that the proportion of false positives is controlled.

In addition, Figure \ref{fig:CASTthreshold}(a) - (b) illustrates that both GBH and GBY methods are more liberal than the LCAST procedure regardless of the magnitude and direction of the average correlations.  Also, Figure \ref{fig:CASTthreshold}(d) shows that the threshold of GBY procedure is more liberal than the proposed QCAST procedure for the uncorrelated case $(\overline{r} = 0)$ and the positively correlated case $(\overline{r} > 0)$. Meanwhile, moderately negatively correlated groups in Figure \ref{fig:CASTthreshold}(c) have a steeper curve that approaches the GBY threshold as the rank of the $P$-value increases. On the average, when there is a strong negative correlation among the variables within a group, the QCAST threshold exceeds the GBH and GBY methods as the order of the $P$-value increases.
\vspace{1.5em}

\section{Results and Discussion}
\label{sec:results}

\subsection{Numerical Experiments}
\label{sec:NumericalStudies}

In this section, we evaluate the performance of CAST procedures using FDR and empirical power vis-\'a-vis the existing grouped BH and BY \citep{benjamini2000adaptive, benjamini2001control}.  Throughout the simulations, we consider the level of significance $\alpha$ at 0.05.

The scientific goal in this study is to identify which of the $M$ probes exhibit differential methylation. Statistically, this translates to $M$ simultaneous tests for a significant difference in the mean methylation level of the $j$th probe in the $g$th gene between cases and controls, for all $j \in [M_g], g \in [G]$, $M_1 + M_2 + \ldots + M_G = M$. Suppose that the total of $N$ subjects can be categorized into $N_0$ controls and $N_1$ cases, with $N_0 + N_1 = N$.  For the simulation settings, the number of observations $N$ is set to 300 because this represents a realistic sample size in terms of recruitment of participants in EWAS \citep{albao2019methylation, langevin2014leukocyte}.  Research protocols generally aim for a balanced allocation of subjects, ensuring an equal number of participants in both case and control groups. However, due to unforeseen factors such as dropout rates or the lower prevalence of rare diseases, the distribution of participants between the case and control groups may become unbalanced.  Hence, the proportion of cases, $\gamma_D$, is set at either 0.25 or 0.50.  Also, the choice of the pairs  $(G, M) \in \{(100, 1250), (9600, 120000), (48000, 600000)\}$ reflects the ratio $M/G$ observed in the high-dimensional and ultra high-dimensional epigenetics data set.

Accordingly, suppose the true mean difference in methylation levels among cases and controls is $\zeta$ and the true variance is $\sigma$.  
To investigate the behavior of the proportion of true and false positives as the effect size increases, the true mean difference is set to $\zeta \in \{0.7,  0.9, 1.1\}$ with a standard deviation of the methylation levels $\sigma = 1$.  Here, $\zeta > 0$ represents the strength of the signals for the non-null case whereas $\zeta = 0$ corresponds to the null setting.  Another factor of interest is the overall proportion of non-nulls $\pi_1 \in \{0.005, 0.01\}$ to introduce sparsity in the identification of significant probes. Likewise, the proportion of groups with at least one significant variable is set to $\psi = 0.05$. 
\vspace{1em}

\begin{figure}[!ht]
    \centering
    \caption{Illustration of the correlation matrix of probes belonging in $G$ genes when the probability of having at least one correlated pair of probes from different genes is $\upsilon$}
    \label{fig:corr-simulated}
    \begin{tabular}{cc}
       \includegraphics[width=0.45\linewidth]{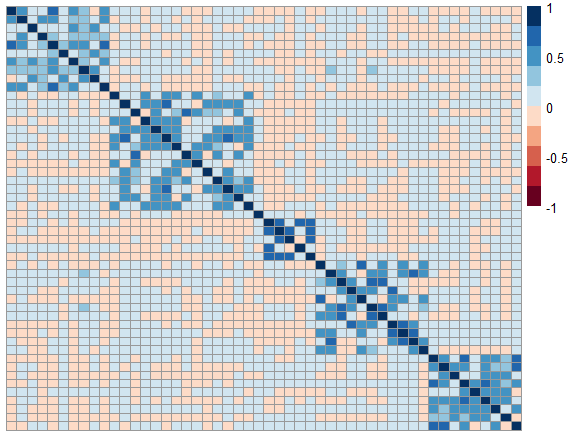} &
       \includegraphics[width=0.45\linewidth]{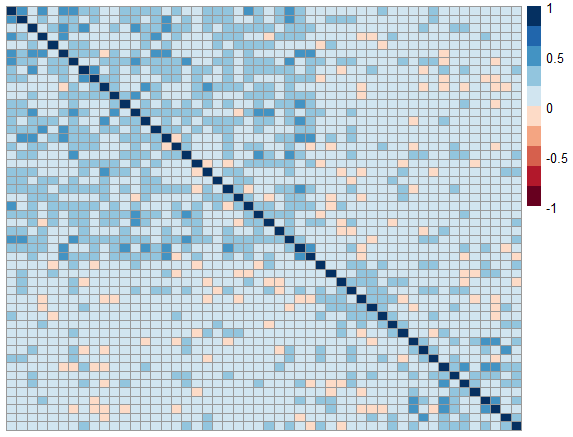}\\
       $(a) ~\upsilon=0, G = 5$ & $(b) ~\upsilon=0.5, G = 3$\\
    \end{tabular}
\end{figure}
\vspace{1em}

Meanwhile, suppose $\kappa^{g}_{jj'}$ is the probability of the $j$th and $j'$th probes belonging in the $g$th gene to be correlated whereas $\tau^{gg'}_{jj'}$ is the probability of probes belonging in different genes to be correlated.  Define $W_g$ as the number of pairs of correlated probes within the $g$th gene while $B_{gg'}$ is the number of pairs of correlated probes between the $g$th and $g'$th gene.  Suppose the probability of having at least one correlated pair of probes from different genes is $\mathbb{P}(B_{gg'} > 0) = \upsilon_{gg'}$. Using $\kappa^{g}_{jj'}, \tau^{gg'}_{jj'}, W_g, B_{gg'}$ and $\upsilon_{gg'}$, the covariance matrix can be constructed to reflect the dependencies of the probes within and between genes. Some adjustments can be implemented to ensure that the covariance matrix to ensure that it is positive definite.  For a given mean vector, the simulated methylation level data is then generated from the multivariate normal distribution.

\begin{figure}[!ht]
    \centering
    \caption{Comparison of the False Discovery Rate among the proposed CAST methods versus GBH and GBY. The visual comparison includes the number of variables $M$, proportion of cases $\gamma_D$, signal strength of the non-null values $\zeta$, and the probability of having at least one correlated pair of probes from different genes, $\upsilon$. }
    \includegraphics[width=\linewidth]{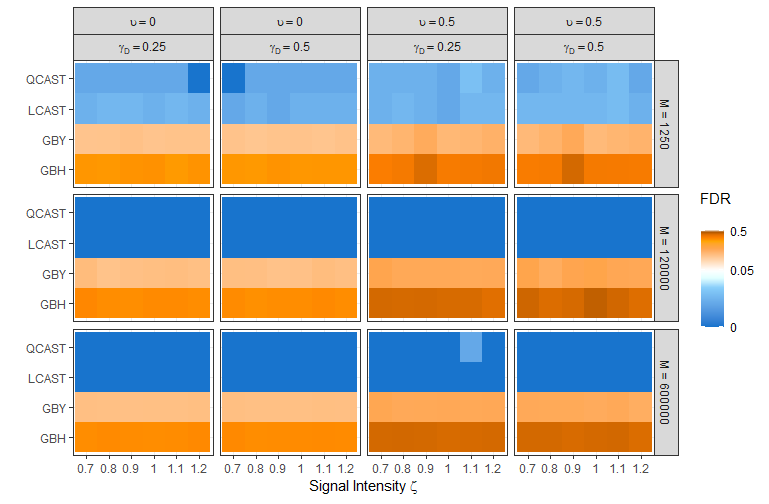} 
    \label{fig:FDR}
\end{figure}

\begin{figure}[!ht]
    \centering
    \caption{Comparison of the Empirical Power among the proposed CAST methods versus GBH and GBY. The visual comparison includes the number of variables $M$, proportion of cases $\gamma_D$, signal strength of the non-null values $\zeta$, and the probability of having at least one correlated pair of probes from different genes, $\upsilon$. The gray line plots indicate that the method failed to control FDR.}
    \includegraphics[width=\linewidth]{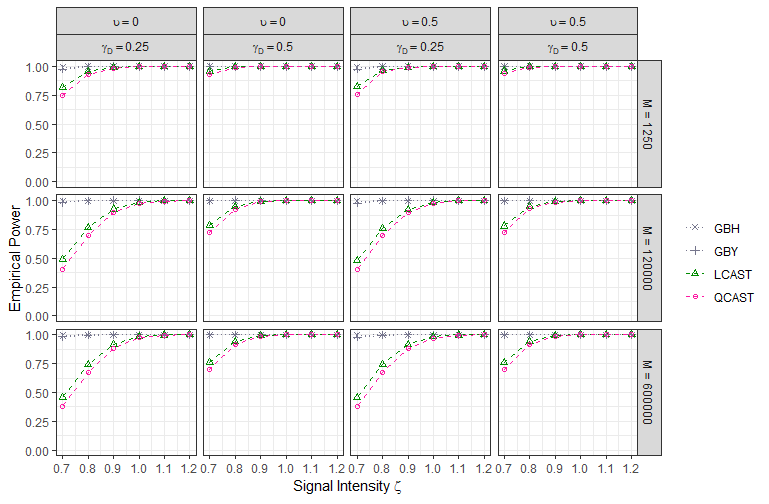} 
    \label{fig:TPR}
\end{figure}

\begin{table}[ht!]
\centering
\label{tab:Simulation1250}
\caption{Numerical comparison of the proposed CAST methods with the grouped version of BH and BY when $M = 1250$. The quantity $\gamma_D$ represents the proportion of cases in the sample while $\zeta$ represents the signal strength of the non-null values. The highlighted results yields the highest true positive rate (TPR) given that the false discovery rate (FDR) is controlled at $\alpha = 0.05$.}
\vspace{1.5em}

\scriptsize{\begin{tabular}{|c|c|ccc|ccc|ccc|ccc|ccc|}
\hline 
$\gamma_D$ & $\zeta$ & \multicolumn{3}{c|}{\textbf{GBH}} & \multicolumn{3}{c|}{\textbf{GBY}} & \multicolumn{3}{c|}{\textbf{LCAST}} & \multicolumn{3}{c|}{\textbf{QCAST}}  
\\ 
\cline{3-14} & & \small $\widehat{\text{R}}$  &\small $\widehat{\text{FDR}}$  & \small $\widehat{\text{TPR}}$  & \small $\widehat{\text{R}}$  &\small $\widehat{\text{FDR}}$  & \small $\widehat{\text{TPR}}$  & \small $\widehat{\text{R}}$  &\small $\widehat{\text{FDR}}$  & \small $\widehat{\text{TPR}}$  & \small $\widehat{\text{R}}$  &\small $\widehat{\text{FDR}}$  & \small $\widehat{\text{TPR}}$ \\ \hline
\multicolumn{14}{|c|}{ $\upsilon=0.5$} \\ \hline

\multirow{6}{*}{0.50} & 0.7 & 21.5 & 0.368 & 1.000 & 15.1 & 0.153 & 0.999 & \cellcolor{yellow}\fontseries{b}\selectfont11.6 & \cellcolor{yellow}\fontseries{b}\selectfont0.002 & \cellcolor{yellow}\fontseries{b}\selectfont0.966 & 11.4 &  0.002 & 0.946\\
& & (10) & (0.173) & (0.004) & (6) & (0.151) & (0.010) & (0.9) & (0.023) & (0.053) & (1.2) & (0.028) & (0.070) \\
& 0.9 & 22.1 & 0.400 & 1.000 & 15.3 & 0.167 & 1.000 & \cellcolor{yellow}\fontseries{b}\selectfont12 & \cellcolor{yellow}\fontseries{b}\selectfont0.003 & \cellcolor{yellow}\fontseries{b}\selectfont0.999 & 12.1 &  0.004 & 0.999 \\ 
& & (7.9) & (0.169) & (0.000) & (4.6) & (0.158) & (0.000) & (0.7) & (0.029) & (0.007) & (0.9) & (0.039) & (0.010) \\
&1.1 & 21.1 & 0.362 & 1.000 & 14.8 & 0.145 & 1.000 & \cellcolor{yellow}\fontseries{b}\selectfont12 & \cellcolor{yellow}\fontseries{b}\selectfont0.002 & \cellcolor{yellow}\fontseries{b}\selectfont1.000 & 12.1 &  0.003 & 1.000 \\ 
& & (8.9) & (0.175) & (0.000) & (4.9) & (0.139) & (0.000) & (0.6) & (0.022) & (0.000) & (0.9) & (0.033) & (0.000) \\
\hline

\multirow{6}{*}{0.25}   & 0.7 & 21.5 & 0.378 & 0.994 & 14.8 & 0.160 & 0.982 & \cellcolor{yellow}\fontseries{b}\selectfont10.1 & \cellcolor{yellow}\fontseries{b}\selectfont0.004 & \cellcolor{yellow}\fontseries{b}\selectfont0.835 & 9.4 &  0.003 & 0.774 \\ 
& & (9.5) & (0.166) & (0.022) & (5.2) & (0.143) & (0.038) & (1.6) & (0.037) & (0.117) & (2) & (0.042) & (0.133) \\
&0.9 & 21.5 & 0.384 & 1.000 & 15.4 & 0.168 & 1.000 & \cellcolor{yellow}\fontseries{b}\selectfont12 & \cellcolor{yellow}\fontseries{b}\selectfont0.003 & \cellcolor{yellow}\fontseries{b}\selectfont0.993 & 11.9 &  0.002 & 0.985 \\
& & (7.7) & (0.169) & (0.004) & (4.9) & (0.163) & (0.006) & (0.8) & (0.028) & (0.024) & (1) & (0.032) & (0.036) \\
& 1.1 & 20.8 & 0.358 & 1.000 & 15 & 0.153 & 1.000 & 12 & 0.002 & 1.000 & \cellcolor{yellow}\fontseries{b}\selectfont12 & \cellcolor{yellow}\fontseries{b}\selectfont 0.001 & \cellcolor{yellow}\fontseries{b}\selectfont 1.000 \\
& & (8.8) & (0.171) & (0.000) & (5.3) & (0.147) & (0.000) & (0.2) & (0.015) & (0.000) & (0.1) & (0.006) & (0.003) \\
\hline

\multicolumn{14}{|c|}{ $\upsilon=0$} \\ \hline
\multirow{6}{*}{0.50}  & 0.7 & 18.2 & 0.325 & 0.999 & 14 & 0.134 & 0.997 & \cellcolor{yellow}\fontseries{b}\selectfont11.5 & \cellcolor{yellow}\fontseries{b}\selectfont0.002 & \cellcolor{yellow}\fontseries{b}\selectfont0.959 & 11.2 & 0.001 & 0.933 \\
& & (2.7) & (0.100) & (0.008) & (1.5) & (0.090) & (0.015) & (0.7) & (0.014) & (0.061) & (0.9) & (0.008) & (0.075) \\
& 0.9 & 18.1 & 0.323 & 1.000 & 14 & 0.134 & 1.000 & 12 & 0.002 & 0.999 & \cellcolor{yellow}\fontseries{b}\selectfont12 & \cellcolor{yellow}\fontseries{b}\selectfont 0.001 & \cellcolor{yellow}\fontseries{b}\selectfont 0.999 \\
& & (2.7) & (0.100) & (0.000) & (1.5) & (0.088) & (0.000) & (0.2) & (0.013) & (0.007) & (0.2) & (0.008) & (0.009) \\
& 1.1 & 17.9 & 0.315 & 1.000 & 14 & 0.135 & 1.000 & 12 & 0.002 & 1.000 & \cellcolor{yellow}\fontseries{b}\selectfont12 & \cellcolor{yellow}\fontseries{b}\selectfont 0.001 & \cellcolor{yellow}\fontseries{b}\selectfont 1.000 \\
& & (2.5) & (0.098) & (0.000) & (1.4) & (0.085) & (0.000) & (0.1) & (0.011) & (0.000) & (0.1) & (0.007) & (0.000) \\
\hline
   
\multirow{6}{*}{0.25} & 0.7 & 18.2 & 0.330 & 0.993 & 13.8 & 0.138 & 0.978 & \cellcolor{yellow}\fontseries{b}\selectfont10.1 & \cellcolor{yellow}\fontseries{b}\selectfont0.003 & \cellcolor{yellow}\fontseries{b}\selectfont0.842 & 9.3 &  0.001 & 0.775 \\
&   & (2.8) & (0.100) & (0.025) & (1.6) & (0.088) & (0.043) & (1.3) & (0.015) & (0.111) & (1.6) & (0.010) & (0.130) \\
& 0.9 & 17.8 & 0.311 & 1.000 & 13.9 & 0.130 & 1.000 & \cellcolor{yellow}\fontseries{b}\selectfont11.9 & \cellcolor{yellow}\fontseries{b}\selectfont0.002 & \cellcolor{yellow}\fontseries{b}\selectfont0.994 & 11.8 &  0.001 & 0.986 \\
& & (2.6) & (0.100) & (0.000) & (1.4) & (0.083) & (0.005) & (0.3) & (0.011) & (0.023) & (0.4) & (0.007) & (0.034) \\ 
& 1.1 & 17.9 & 0.316 & 1.000 & 13.9 & 0.128 & 1.000 & 12 & 0.001 & 1.000 & \cellcolor{yellow}\fontseries{b}\selectfont12 & \cellcolor{yellow}\fontseries{b}\selectfont 0.000 & \cellcolor{yellow}\fontseries{b}\selectfont 1.000 \\
& & (2.6) & (0.101) & (0.000) & (1.4) & (0.086) & (0.000) & (0.1) & (0.011) & (0.003) & (0.1) & (0.006) & (0.005) \\
\hline
\end{tabular}}
\end{table}

An illustration of the correlation matrix for a given $G$ when $\upsilon_{gg'} = \upsilon$ is shown in Figure \ref{fig:corr-simulated}.  For the simulation settings, the case of independent groups is considered when $\upsilon = 0$ which is close to the block-correlation setting of \cite{stevens2017comparison} while we set the
probability of having at least one correlated pair of probes from different genes to be $\upsilon_{gg'} = \upsilon = 0.5$ for any $g, g' \in [G]$.  For the latter case, we set the probability of probes belonging in different genes to be correlated as $\tau^{gg'}_{jj'} = \tau = 0.5$, for some $g \neq g'$ and $j \neq j'$.  In reality, not all of the probes within the same group will be correlated. As such, we fixed the probability of probes in the same group to be correlated, i.e., $\kappa^{g}_{jj'} = \kappa = 0.5$. In Figure \ref{fig:corr-simulated}(a) when $\upsilon_{gg'} = \upsilon = 0$, this implies that none of the $M_g$ probes in the $g$th gene and $M_{g'}$ probes in the $g'$th gene are correlated, for any $g, g' = 1, 2, \ldots, 5$. In contrast, Figure \ref{fig:corr-simulated}(b) presents the case where there is correlation among probes within the same gene and probes from different genes are also correlated.

Figure \ref{fig:FDR} compares the False Discovery Rate (FDR) among the CAST methods, GBH, and GBY. The results clearly show that the CAST procedures effectively control the FDR across all scenarios, whereas GBH and GBY exhibit inflated FDRs. Additionally, GBH and GBY demonstrate a higher proportion of false discoveries when there is correlation across groups, compared to scenarios where groups are independent. In contrast, differences in FDR among the procedures are minimal with respect to within-group correlation and the balance of case-control counts in the sample.

Consequently, Figure \ref{fig:TPR} displays the comparison of empirical power among the CAST methods, GBH, and GBY. 
Likewise, we show the numerical comparison when $M = 1250$ in Table \ref{tab:Simulation1250}.  These tabular values corresponds to the first row illustrated in Figure \ref{fig:FDR} and \ref{fig:TPR}.
The results indicate that the proposed CAST procedures achieved desirable empirical power. Both CAST methods demonstrated a consistent pattern of increasing empirical power with higher signal intensity. LCAST showed slightly higher empirical power than QCAST, particularly for weaker signal intensities. However, both LCAST and QCAST experienced a slight reduction in power when the sampling of cases and controls was highly unbalanced. Other factors, such as signal sparsity, correlation between groups, and within-group correlation, did not produce significant differences in the empirical power of the procedures.
\vspace{1.5em}

\subsection{Analysis of blood-based EWAS to identify significant differentially methylated CpG loci associated with bladder cancer}
\label{sec:DNAMethylation}

According to the American Cancer Society \cite{acs2024bladder}, approximately 83,000 individuals are diagnosed with bladder cancer each year in the United States, and approximately 430,000 patients worldwide, resulting in over 165,000 deaths annually \citep{ferlay2013cancer}. This type of cancer predominantly affects older adults, with nearly 90\% of cases occurring in people aged 55 and older, and the average age at diagnosis being 73.  Despite its prevalence, bladder cancer is frequently mismanaged \citep{kamat2016bladder}. A 2012 analysis of data from the Surveillance, Epidemiology, and End Results Program revealed that among 4,790 patients diagnosed with high-grade non-muscle-invasive bladder cancer between 1992 and 2002, only one received treatment in line with established guidelines \citep{chamie2012quality}.  In addition, bladder cancer is typically incurable in the metastatic stage, and no systemic therapy has proven effective in extending survival for patients who have advanced after first-line cisplatin-based chemotherapy \citep{iyer2013prevalence}.  To enhance adherence to best practices in bladder cancer treatment, there is an urgent need for novel approaches, including a comprehensive review of current diagnostic and management strategies for this disease.

With the goal of contributing to the advancement of precision medicine-based approach for patients with bladder cancer, we developed a novel set of analytical methods to identify significant bladder cancer-associated CpG loci in blood, accounting for leukocyte composition.  We analyzed an epigenome-wide association study (EWAS) on blood specimens from a case-control study of bladder cancer by \cite{langevin2014leukocyte} (GSE50409).  They used Illumina Infinium 27k Human DNA Methylation Beadchip v1.2 to obtain DNA methylation profiles of peripheral blood samples from 223 newly diagnosed bladder cancer cases and 205 healthy controls with no prior history of cancer.  After adjusting for the false discovery rate, \cite{langevin2014leukocyte} reported 3,987 differentially methylated loci $(\text{adjusted $P$-value} \leq 0.05)$ for bladder cancer cases compared to controls.

On the other hand, when using genes as the grouping variable for the CpG loci, the grouped Benjamini-Hochberg procedure in \eqref{eqn:GBH} resulted in a total of 2,784 differentially methylated probes (DMPs), with an $\text{adjusted $P$-value} \leq 0.05$.  Likewise, the grouped Benjamini-Yekutieli method in \eqref{eqn:GBY} yielded 2,190 DMPs $(\text{adjusted $P$-value} \leq 0.05)$. In contrast, the proposed methods LCAST and QCAST resulted to 20 and 18 DMPs as shown in Table \ref{tab:DMPs}.  Our numerical studies confirmed that existing grouped BH and BY procedures consistently fail to control the false discovery rate (FDR) across all scenarios, particularly when group sizes are small, such as when only a few methylation probes are associated with a given gene.  Additionally, our simulations demonstrated that both LCAST and QCAST effectively controlled the FDR while maintaining consistent empirical power across all scenarios, achieving a true discovery rate comparable to the grouped BH and BY procedures.

\begin{table}[t!]
\centering
\caption{Resulting Differentially Methylated CpG Loci using CAST procedures when $\alpha = 0.05$. Highlighted values are significant using the CAST methods but are not found to be significant in both the main and supplementary text of \cite{langevin2014leukocyte}.}
\label{tab:DMPs}
\vspace{1.5em}
\begin{tabular}{|cc|cc|c|c|}
\hline
&& \multicolumn{2}{c|}{Adjusted $P$-values} & \multicolumn{2}{c|}{DMPs found in}\\
\cline{3-6}  CpG Loci&Gene	&LCAST	&QCAST	&Main &	Supp\\
\hline
cg25307081	&BRD7	&0.000045	&0.000057	&Yes &Yes\\
cg07510052	&HIGD2A	&0.000603	&0.000795	&Yes &Yes\\
cg20150565	&ZNFX1	&0.003385	&0.004231	&Yes &Yes\\
cg13064571	&C8orf44 &0.003422  &0.003422	&No	 &Yes\\
cg00429618	&ZNF322B&0.003480	&0.004509	&Yes &Yes\\
cg12466095	&C14orf103 &0.005337 &0.005337	&No	 &Yes\\
cg07081888	&NALP4	&0.005401	&0.006751	&Yes &Yes\\
cg16729794	&MOBP	&0.008406   &0.008406   &No	 &Yes\\
cg26413827	&DACH1	&0.012236	&0.015481	&No	 &Yes\\
cg00410898  &STC1   &0.014526   &0.0145262  &\cellcolor{pink}\fontseries{b}\selectfont No &\cellcolor{pink}\fontseries{b}\selectfont No \\
cg08578023	&CTSS	&0.018235	&0.018235	&\cellcolor{pink}\fontseries{b}\selectfont No &\cellcolor{pink}\fontseries{b}\selectfont No \\
cg00406188	&LCE2C	&0.024811	&0.032357	&No	 &Yes\\
cg18156583	&IL18RAP&0.026862	&0.034752	&  \cellcolor{pink}\fontseries{b}\selectfont No	&  \cellcolor{pink}\fontseries{b}\selectfont No\\
cg04070847	&CD200R1&0.032020	&0.040467	&No	&Yes\\
cg09425215	&CHD2	&0.035473	&0.051442	&Yes &Yes\\
cg08972170	&Ell1	&0.035590	&0.044487	&  \cellcolor{pink}\fontseries{b}\selectfont No	&  \cellcolor{pink}\fontseries{b}\selectfont No\\
cg10789261	&LARP5	&0.036247	&0.036247 &No &Yes\\
cg02245418	&ZNF364	&0.039724	&0.039724  &No &Yes\\
cg16468729	&IL8	&0.044508	&0.055635	&  \cellcolor{pink}\fontseries{b}\selectfont No	&  \cellcolor{pink}\fontseries{b}\selectfont No\\
cg10671066	&SLAMF6	&0.046206 &0.046206	&\cellcolor{pink}\fontseries{b}\selectfont No &\cellcolor{pink}\fontseries{b}\selectfont No \\
\hline
\end{tabular}
\end{table}
As illustrated in Table \ref{tab:DMPs}, the CAST methods validate the findings from \cite{langevin2014leukocyte}, where differential methylation between bladder cancer cases and controls was detected at CpG sites within the BRD7, HIGD2A, ZNFX1, C8orf44, ZNF322B, C14orf103, NALP4, MOBP, DACH1, LCE2C, CD200R1, CHD2, LARP5, and ZNF364 genes.  To demonstrate the practical applicability of the findings using the CAST methods, we investigate how some of these genes can guide researchers in formulating new hypotheses and in the development of novel therapeutic targets.

BRD7 is extensively expressed across various human tissues, including the brain, heart, lung, colon, and breast \citep{peng2002analysis}. Numerous studies suggest that BRD7 interacts with a variety of transcription factors and plays crucial roles in diverse cellular processes, such as carcinogenesis, chromatin remodeling, cell proliferation, apoptosis, glucose metabolism, and transcriptional regulation \citep{yu2016brd7, kim2016role, cong2006transcriptional, zhou2004brd7, park2020emerging, zhou2006identification}. A growing body of evidence indicates that BRD7 is downregulated in several cancer types, including nasopharyngeal carcinoma \citep{peng2007brd7, liu2016mir, yu2000analysis, li2023brd7}, ovarian cancer \citep{park2014tumor}, gastric cancer \citep{zhang2003expression}, colorectal cancer \citep{wu2013prognostic}, breast cancer \citep{niu2020brd7}, and prostate cancer \citep{kikuchi2009trim24}. The involvement of BRD7 in multiple cancers suggests its potential as a target for anti-cancer therapies, where BRD7 could serve as a molecular biomarker for cancer diagnosis \citep{yu2016brd7}. Additionally, BRD7 has been identified as a tumor suppressor gene (TSG) and nuclear transcriptional regulator (NTR) in various malignancies \citep{liu2018brd7, penkert2012no}. Monitoring BRD7 levels, including gene and protein alterations, could help predict cancer risk, progression, and prognosis \citep{yu2016brd7}. Given its reduced expression in many cancers, strategies to reintroduce the BRD7 gene or stabilize BRD7 protein levels could be promising approaches for cancer management. Further researches involving restoring its tumor-suppressing function may inhibit tumor growth, offering a potential target for cancer therapy.

Moreover, the Hypoxia Inducible Gene Domain Family Member 2A (HIGD2A) protein is crucial for the assembly of the mitochondrial respiratory supercomplex, which plays a significant role in promoting cell proliferation and survival under hypoxic conditions \citep{salazar2020biosystem, huang2023higd2a}. High-resolution expression profiling indicates that HIGD2A is involved in cancer biology \citep{edgar2002gene}. While HIGD2A is not classified as a cancer-driving gene \citep{sondka2018cosmic}, significant alterations in its DNA methylation and mRNA expression have been detected in various cancers. A comparison of average beta values between tumor and matched normal samples reveals substantial changes in DNA methylation and mRNA expression of the HIGD2A gene in several cancers, including but not limited to breast, liver, lung, pancreatic, and prostate cancers \citep{salazar2020biosystem}. Specifically, knockdown of HIGD2A was shown to inhibit the proliferation and growth of hepatocellular carcinoma (HCC) cells by disrupting the MAPK/ERK pathway, suggesting a role for HIGD2A in abnormal mitochondrial metabolism in HCC \citep{huang2023higd2a}. Further research should aim to clarify the precise mechanism of HIGD2A's involvement in modulating the MAPK/ERK pathway, investigate its role in metastasis, and explore its potential as a target for cancer therapy.

As the largest family of transcription factors in the human genome, zinc finger proteins, with their diverse combinations and functions, play critical roles in various biological processes, including those related to tumorigenesis and tumor progression \citep{jen2016zinc, li2022structures}. ZNF322 is extensively expressed in human tissues during both embryonic and adult stages, with its protein localized in both the nucleus and cytoplasm. The overexpression of ZNF322 suggests that it may act as a positive transcriptional regulator in MAPK-mediated signaling pathways \citep{li2004znf322}. Furthermore, the overexpression of the zinc finger protein, X-linked (ZFX), has been associated with promoting cell growth and metastasis in a variety of cancers, including hepatocellular carcinoma \citep{lai2014overexpression}, nasopharyngeal carcinoma \citep{li2015high}, non-small cell lung cancer \citep{jiang2012role}, gastric cancer \citep{wu2013knockdown}, gallbladder cancer \citep{weng2015zinc}, and breast cancer \citep{yang2014shrna}. Additionally, Zinc Finger Antisense 1, also known as ZNFX1 Antisense RNA1, was originally identified as a tumor suppressor gene in breast cancer \citep{askarian2011snord} and liver cancer \citep{wang2016long}.  Inhibiting ZFX using siRNA oligos or drug treatments has been shown to suppress cancer progression, indicating the potential of oncogenic zinc finger proteins as therapeutic targets \citep{li2013zfx, liu2015baicalein}.

Meanwhile, NALPs belong to a large superfamily of proteins that are integral to innate immunity, known as NOD-like receptors (NLRs) \citep{martinon2005nlrs}. NALP1, in particular, acts as a scaffolding protein within the inflammasome complex, where it has been demonstrated to activate caspase-1 and facilitate the maturation of the endogenous pyrogen, interleukin-1$\beta$ (IL-1$\beta$) 
\citep{martinon2002inflammasome}. Also, multiple studies have identified mutations in the NALP1-related gene, NALP3, as the cause of various periodic fever syndromes, such as Muckle-Wells syndrome (MWS), familial cold autoinflammatory syndrome (FCAS), and chronic infantile neurological cutaneous and articular (CINCA) syndrome, also known as neonatal-onset multisystem inflammatory disease \citep{church2006hereditary, hoffman2001mutation}. However, the function of NALP4 has not been as extensively explored as that of other members of the NALP subfamily, like NALP1, NALP3, and NALP5 \citep{tschopp2003nalps}. The human genome comprises 14 NALP family members, suggesting that other NALPs may also form inflammasomes and participate in immune responses \citep{tschopp2003nalps, martinon2007nalp}. Beyond their role in the activation of inflammatory cytokines, some research has identified NALPs as potential candidate genes involved in cancer \citep{sjoblom2006consensus}.  Given the limited understanding of NALP4, future studies and novel mouse models are expected to provide deeper insights into the roles of various NALP4 proteins in human infections, inflammatory diseases, and possibly uncover or confirm new roles for NALP4 in other biological pathways and conditions.

DACH1, an essential component of the retinal determination gene network (RDGN) family, has been increasingly recognized for its diverse roles in tumorigenesis and metastasis \citep{wu2015six1, popov2010dachshund}. Multiple studies have highlighted the potential of DACH1 as a prognostic marker in breast cancer, noting a significant correlation between lower DACH1 expression and poor clinical outcomes \citep{wu2006dach1, powe2014dach1, wu2014cell}. There is also considerable attention directed towards DACH1's ability to inhibit Epithelial-Mesenchymal Transition (EMT) and diminish the population of cancer stem cells (CSCs) \citep{wu2014cell, wu2011cell}, supporting its role as a novel tumor suppressor. Furthermore, the epigenetic silencing of DACH1 in various cancers has been shown to be reversible with histone deacetylase inhibitors, leading to reduced cancer cell proliferation \citep{zhu2013epigenetic, yan2013epigenetic}. This suggests a potential therapeutic approach using demethylase treatment to restore DACH1 expression in patients with hepatocellular carcinoma \citep{liu2015dach1}.

In addition, the Late Cornified Envelope (LCE) gene families are predominantly expressed in the epidermis \citep{jackson2005late} and play roles in epidermal growth, peptide cross-linking, keratinocyte differentiation, and keratinization \citep{habib2022differential}. Several studies have noted the downregulation of LCE2C in conditions such as periodontitis \citep{jeon2021reliability} and tongue squamous cell carcinoma (TSCC) \citep{boldrup2017gene}. Additionally, aberrant expression of LCE2C has been observed in oral squamous cell carcinoma (OSCC) \citep{zhang2021identification} and head and neck squamous cell carcinoma (HNSCC) \citep{liang2020mmp25}.  A transcriptomic analysis examining the cross-talk between periodontitis and hypothyroidism identified LCE2C as one of the top 10 genes with the most connections \citep{yan2022transcriptomic}. Further investigation into LCE2C could uncover its role as a potential crosstalk gene between various diseases.

CD200, a member of the immunoglobulin superfamily, interacts with its receptor CD200R1 to influence the immune microenvironment within cancers \citep{yoshimura2020cd200}. CD200 is also expressed in various organs, including the skin and central nervous system, as well as in tumor cells \citep{broderick2002constitutive}. \cite{coles2012expression} demonstrated that CD200 on leukemia cells can directly suppress T-cell responses, highlighting the potential of CD200-blocking therapy in treating acute myeloid leukemia (AML). Both experimental and clinical evidence show that blocking CD200/CD200R1 signaling can enhance Th1-related cytokine expression and inflammation, thereby promoting anticancer responses \citep{siva2008immune}. As a result, the CD200/CD200R1 pathway is regarded as a promising therapeutic target \citep{yoshimura2020cd200}.

Like BRD7, DACH1, and ZNFX1 Antisense RNA1, the CHD2 protein functions as a tumor suppressor and may play a significant role in modulating DNA damage responses at the chromatin level \citep{nagarajan2009role}. The mutation of CHD2 in mice results in lethality, highlighting its critical importance in embryonic development \citep{nagarajan2009role}. Experiments on CHD2 using mice exhibit an increased susceptibility to lymphoma and abnormalities in their hematopoietic system, where stem cells undergo abnormal differentiation \citep{nagarajan2009role}. Additionally, \cite{nagarajan2009role} revealed that mouse embryonic fibroblasts derived from CHD2 heterozygotes showed a higher mutation rate, prompting further investigation into CHD2's role in the DNA damage response \citep{stanley2013chd}.

Consequently, LCAST and QCAST not only validate results from \cite{langevin2014leukocyte}.  More importantly, these methods also uncover novel results at CpG sites within the STC1, CTSS, IL18RAP, Ell1, IL8, and SLAMF6 genes that were not previously identified by \cite{langevin2014leukocyte}, offering new avenues for further exploration. In particular, polymorphisms in the IL18RAP region are linked to several immune-mediated diseases, including inflammatory bowel disease (IBD) \citep{zhernakova2008genetic}, atopic dermatitis \citep{hirota2012genome}, leprosy \citep{liu2012identification}, celiac disease \citep{hunt2008newly}, and type I diabetes \citep{smyth2008shared}. The interaction between IL18RAP and IL-18R1 is crucial for the signal transduction initiated by IL-18 \citep{boraschi200618}. While IL-18 signaling is well recognized for mediating Th1 responses \citep{boraschi200618}, it also plays a role in various biological processes, such as maintaining the integrity of the intestinal epithelial barrier and responding to commensal microbiota \citep{elinav2011nlrp6, huber201222bp, zaki2010nlrp3}. Reduced IL-18 induction, due to impaired inflammasome activity, can worsen experimental colitis and intestinal injury \citep{elinav2011nlrp6, huber201222bp}. Additionally, IL-18 is involved in resolving lung infections \citep{kawakami199718}. However, IL-18 administration has been shown to induce murine colitis \citep{chikano200018} and lupus-like disease \citep{esfandiari2001proinflammatory}. Therefore, it is critical to balance IL-18 signaling, as it can have both protective and harmful effects. The impact of IL18RAP polymorphisms on pathways initiated by multiple pattern-recognition receptors (PRRs) in monocyte-derived macrophages (MDMs) suggests that the IL-18 pathway could be a significant therapeutic target in immune-mediated diseases. These findings highlight the need for caution when designing therapies targeting the IL-18 pathway for inflammatory diseases, as enhancing or maintaining, rather than inhibiting, IL-18 signaling may be more beneficial in certain conditions \citep{hedl2014il18rap}.

On the other hand, the Eleven-Nineteen Lysine-Rich Leukemia (ELL) family of proteins represents a newly identified class of transcription elongation factors (TEFs) \citep{sweta2021functional}. ELL1 was initially discovered as a fusion partner of the mixed lineage leukemia (MLL) protein in acute myeloid leukemia \citep{thirman1994cloning}. The deletion of ELL leads to embryonic lethality in organisms such as \textit{C. elegans, Drosophila,} and mice \citep{cai2011regulation, eissenberg2002dell, mitani2000nonredundant}. However, the roles and mechanisms of action of ELL across different organisms are complex and require further detailed investigation.

Interleukin-8 (IL-8) expression is regulated by various stimuli, including inflammatory signals (e.g., tumor necrosis factor-$\alpha$, IL-1$\beta$), chemical and environmental stresses (e.g., chemotherapy agents and hypoxia), and steroid hormones (e.g., androgens, estrogens, and dexamethasone) \citep{brat2005role}. Additionally, IL-8 is a potent chemoattractant for neutrophils, which act as ``early responders'' to wounds and infections. These neutrophils release enzymes that remodel the extracellular matrix of the tissues they traverse to reach the site of injury or infection \citep{nathan2006neutrophils}. \cite{de2004potential} proposed that the cellular response to IL-8 released by tumor cells enhances angiogenesis, thereby contributing to tumor growth and progression. The enzymes released by neutrophils could facilitate tumor cell migration through the extracellular matrix, aiding in their entry into the vasculature and subsequent metastasis.  The presence of IL-8 receptors on cancer cells, endothelial cells, neutrophils, and tumor-associated macrophages indicates that IL-8 secretion from cancer cells can significantly impact the tumor microenvironment \citep{waugh2008interleukin}. Activation of IL-8 receptors on endothelial cells triggers several signaling pathways, promoting an angiogenic response by inducing proliferation, survival, and migration of vascular endothelial cells \citep{brat2005role, li20038}.  Moreover, as observed in numerous clinical studies, targeting IL-8 or its associated chemokines could have significant implications for the systemic treatment of aggressive and metastatic diseases \citep{waugh2008interleukin}.

Through the application of our proposed methods, we have demonstrated that DNA methylation in peripheral blood holds potential as a biomarker for assessing bladder cancer risk. Although many of the genes linked to the differentially methylated CpG sites are not traditionally associated with bladder cancer, their roles and mechanisms in disease progression present promising targets for cancer therapy. Further investigation into the epigenetic regulation of these genes may provide critical insights into their functions in cancer biology and therapeutic resistance. Notably, the findings from the CAST methods significantly reduce the need to analyze tens of thousands of CpG loci and genes as described in \cite{langevin2014leukocyte}, thereby offering substantial savings in time, labor, and financial resources required for clinical studies, animal models, and future therapeutic research.
\vspace{1.5em}

\section{Conclusions and Future Work}
\label{sec:conclusion}

Driven by the insights gained from epigenome-wide association studies, we developed the innovative CAST methods, which are data-adaptive  and correlation-adjusted multiple testing method that effectively control the False Discovery Rate (FDR) while optimizing power for ultra high-dimensional data characterized by sparse signals and inherent grouping structures. This novel class of correlation-adjusted methods accounts for dependencies among variables within the same group, as well as between groups, ensuring that the proportion of false positives remains within the desired threshold.  LCAST is generally prescribed for ultra high-dimensional correlated datasets, even in the presence of substantial variability among group sizes. Our results suggest that the findings from the CAST methods have the potential to drive groundbreaking biological discoveries.

Among the many possible extensions of this work involve overlapping groups or variables belonging in multiple groups since this study only considered methylation probes belonging in mutually exclusive genes. This may be the case when we consider the biological pathways as groups because genes may belong in multiple pathways. Potentially, we can gain relevant biomedical insights from this type of analysis that can better explain how therapeutic tools can be later formulated.

On the clinical aspect, methylation profiling of genes should be established in bladder cancer (BC) to enhance diagnostic sensitivity and specificity. Validation of these methylation markers in clinical BC specimens is necessary for their practical use \citep{enokida2008epigenetics}. Additionally, further studies should prioritize genome-wide screening to identify new methylated genes specific to BC using advanced high-throughput techniques like CpG island microarray assays \citep{aleman2008identification}.

\bibliographystyle{apalike}
\bibliography{arXiV/arXiV_references}

\begin{thebibliography}{}

\bibitem[Albao et~al., 2019]{albao2019methylation}
Albao, D.~S., Cutiongco-de~la Paz, E.~M., Mercado, M.~E., Lirio, A., Mariano, M., Kim, S., Yangco, A., Melegrito, J., Wad-Asen, K., Gauran, I.~I., et~al. (2019).
\newblock Methylation changes in the peripheral blood of {F}ilipinos with {T}ype 2 {D}iabetes suggest spurious transcription initiation at {TXNIP}.
\newblock {\em Human {M}olecular {G}enetics}, 28(24):4208--4218.

\bibitem[Aleman et~al., 2008]{aleman2008identification}
Aleman, A., Adrien, L., Lopez-Serra, L., Cordon-Cardo, C., Esteller, M., Belbin, T., and Sanchez-Carbayo, M. (2008).
\newblock Identification of dna hypermethylation of sox9 in association with bladder cancer progression using cpg microarrays.
\newblock {\em British Journal of Cancer}, 98(2):466--473.

\bibitem[Anastasiadi et~al., 2018]{anastasiadi2018consistent}
Anastasiadi, D., Esteve-Codina, A., and Piferrer, F. (2018).
\newblock Consistent inverse correlation between {DNA} methylation of the first intron and gene expression across tissues and species.
\newblock {\em Epigenetics \& {C}hromatin}, 11(1):1--17.

\bibitem[Askarian-Amiri et~al., 2011]{askarian2011snord}
Askarian-Amiri, M.~E., Crawford, J., French, J.~D., Smart, C.~E., Smith, M.~A., Clark, M.~B., Ru, K., Mercer, T.~R., Thompson, E.~R., Lakhani, S.~R., et~al. (2011).
\newblock Snord-host rna zfas1 is a regulator of mammary development and a potential marker for breast cancer.
\newblock {\em Rna}, 17(5):878--891.

\bibitem[Battagli et~al., 2003]{battagli2003promoter}
Battagli, C., Uzzo, R.~G., Dulaimi, E., Ibanez~de Caceres, I., Krassenstein, R., Al-Saleem, T., Greenberg, R.~E., and Cairns, P. (2003).
\newblock Promoter hypermethylation of tumor suppressor genes in urine from kidney cancer patients.
\newblock {\em Cancer research}, 63(24):8695--8699.

\bibitem[Baylin and Jones, 2016]{baylin2016epigenetic}
Baylin, S.~B. and Jones, P.~A. (2016).
\newblock Epigenetic determinants of cancer.
\newblock {\em Cold Spring Harbor perspectives in biology}, 8(9):a019505.

\bibitem[Benjamini and Hochberg, 1995]{benjamini1995controlling}
Benjamini, Y. and Hochberg, Y. (1995).
\newblock Controlling the false discovery rate: a practical and powerful approach to multiple testing.
\newblock {\em Journal of the Royal {S}tatistical {S}ociety: {S}eries {B} (Methodological)}, 57(1):289--300.

\bibitem[Benjamini and Hochberg, 2000]{benjamini2000adaptive}
Benjamini, Y. and Hochberg, Y. (2000).
\newblock On the adaptive control of the false discovery rate in multiple testing with independent statistics.
\newblock {\em Journal of {E}ducational and {B}ehavioral {S}tatistics}, 25(1):60--83.

\bibitem[Benjamini et~al., 2006]{benjamini2006adaptive}
Benjamini, Y., Krieger, A.~M., and Yekutieli, D. (2006).
\newblock Adaptive linear step-up procedures that control the false discovery rate.
\newblock {\em Biometrika}, 93(3):491--507.

\bibitem[Benjamini and Yekutieli, 2001]{benjamini2001control}
Benjamini, Y. and Yekutieli, D. (2001).
\newblock The control of the false discovery rate in multiple testing under dependency.
\newblock {\em The {A}nnals of {S}tatistics}, 29(4):1165--1188.

\bibitem[Bergman and Cedar, 2013]{bergman2013dna}
Bergman, Y. and Cedar, H. (2013).
\newblock {DNA} methylation dynamics in health and disease.
\newblock {\em Nature {S}tructural \& {M}olecular {B}iology}, 20(3):274--281.

\bibitem[Boldrup et~al., 2017]{boldrup2017gene}
Boldrup, L., Gu, X., Coates, P.~J., Norberg-Spaak, L., Fahraeus, R., Laurell, G., Wilms, T., and Nylander, K. (2017).
\newblock Gene expression changes in tumor free tongue tissue adjacent to tongue squamous cell carcinoma.
\newblock {\em Oncotarget}, 8(12):19389.

\bibitem[Boraschi and Dinarello, 2006]{boraschi200618}
Boraschi, D. and Dinarello, C.~A. (2006).
\newblock Il-18 in autoimmunity.
\newblock {\em European cytokine network}, 17(4):224--252.

\bibitem[Brat et~al., 2005]{brat2005role}
Brat, D.~J., Bellail, A.~C., and Van~Meir, E.~G. (2005).
\newblock The role of interleukin-8 and its receptors in gliomagenesis and tumoral angiogenesis.
\newblock {\em Neuro-oncology}, 7(2):122--133.

\bibitem[Broderick et~al., 2002]{broderick2002constitutive}
Broderick, C., Hoek, R.~M., Forrester, J.~V., Liversidge, J., Sedgwick, J.~D., and Dick, A.~D. (2002).
\newblock Constitutive retinal cd200 expression regulates resident microglia and activation state of inflammatory cells during experimental autoimmune uveoretinitis.
\newblock {\em The American journal of pathology}, 161(5):1669--1677.

\bibitem[Cai et~al., 2011]{cai2011regulation}
Cai, L., Phong, B.~L., Fisher, A.~L., and Wang, Z. (2011).
\newblock Regulation of fertility, survival, and cuticle collagen function by the caenorhabditis elegans eaf-1 and ell-1 genes.
\newblock {\em Journal of Biological Chemistry}, 286(41):35915--35921.

\bibitem[Chamie et~al., 2012]{chamie2012quality}
Chamie, K., Saigal, C.~S., Lai, J., Hanley, J.~M., Setodji, C.~M., Konety, B.~R., Litwin, M.~S., and in~America~Project, U.~D. (2012).
\newblock Quality of care in patients with bladder cancer: a case report?
\newblock {\em Cancer}, 118(5):1412--1421.

\bibitem[Chikano et~al., 2000]{chikano200018}
Chikano, S., Sawada, K., Shimoyama, T., Kashiwamura, S., Sugihara, A., Sekikawa, K., Terada, N., Nakanishi, K., and Okamura, H. (2000).
\newblock Il-18 and il-12 induce intestinal inflammation and fatty liver in mice in an ifn-$\gamma$ dependent manner.
\newblock {\em Gut}, 47(6):779--786.

\bibitem[Church et~al., 2006]{church2006hereditary}
Church, L.~D., Churchman, S.~M., Hawkins, P.~N., and McDermott, M.~F. (2006).
\newblock Hereditary auto-inflammatory disorders and biologics.
\newblock In {\em Springer seminars in immunopathology}, volume~27, pages 494--508. Springer.

\bibitem[Coles et~al., 2012]{coles2012expression}
Coles, S., Hills, R.~K., Wang, E. C.~Y., Burnett, A.~K., Man, S., Darley, R.~L., and Tonks, A. (2012).
\newblock Expression of cd200 on aml blasts directly suppresses memory t-cell function.
\newblock {\em Leukemia}, 26(9):2148--2151.

\bibitem[Cong et~al., 2006]{cong2006transcriptional}
Cong, P., Jie, Z., Ying, L.~H., Ming, Z., Li, W.~L., Hong, Z.~Q., Xin, Y.~Y., Wei, X., Rong, S.~S., Ling, L.~X., et~al. (2006).
\newblock The transcriptional regulation role of brd7 by binding to acetylated histone through bromodomain.
\newblock {\em Journal of cellular biochemistry}, 97(4):882--892.

\bibitem[Costello et~al., 2000]{costello2000aberrant}
Costello, J.~F., Plass, C., and Cavenee, W.~K. (2000).
\newblock Aberrant methylation of genes in low-grade astrocytomas.
\newblock {\em Brain tumor pathology}, 17:49--56.

\bibitem[Dawson and Kouzarides, 2012]{dawson2012cancer}
Dawson, M.~A. and Kouzarides, T. (2012).
\newblock Cancer epigenetics: from mechanism to therapy.
\newblock {\em cell}, 150(1):12--27.

\bibitem[De~Larco et~al., 2004]{de2004potential}
De~Larco, J.~E., Wuertz, B.~R., and Furcht, L.~T. (2004).
\newblock The potential role of neutrophils in promoting the metastatic phenotype of tumors releasing interleukin-8.
\newblock {\em Clinical Cancer Research}, 10(15):4895--4900.

\bibitem[Edgar et~al., 2002]{edgar2002gene}
Edgar, R., Domrachev, M., and Lash, A.~E. (2002).
\newblock Gene expression omnibus: Ncbi gene expression and hybridization array data repository.
\newblock {\em Nucleic acids research}, 30(1):207--210.

\bibitem[Eissenberg et~al., 2002]{eissenberg2002dell}
Eissenberg, J.~C., Ma, J., Gerber, M.~A., Christensen, A., Kennison, J.~A., and Shilatifard, A. (2002).
\newblock dell is an essential rna polymerase ii elongation factor with a general role in development.
\newblock {\em Proceedings of the National Academy of Sciences}, 99(15):9894--9899.

\bibitem[Elinav et~al., 2011]{elinav2011nlrp6}
Elinav, E., Strowig, T., Kau, A.~L., Henao-Mejia, J., Thaiss, C.~A., Booth, C.~J., Peaper, D.~R., Bertin, J., Eisenbarth, S.~C., Gordon, J.~I., et~al. (2011).
\newblock Nlrp6 inflammasome regulates colonic microbial ecology and risk for colitis.
\newblock {\em Cell}, 145(5):745--757.

\bibitem[Enokida and Nakagawa, 2008]{enokida2008epigenetics}
Enokida, H. and Nakagawa, M. (2008).
\newblock Epigenetics in bladder cancer.
\newblock {\em International journal of clinical oncology}, 13:298--307.

\bibitem[Esfandiari et~al., 2001]{esfandiari2001proinflammatory}
Esfandiari, E., McInnes, I.~B., Lindop, G., Huang, F.-P., Field, M., Komai-Koma, M., Wei, X.-q., and Liew, F.~Y. (2001).
\newblock A proinflammatory role of il-18 in the development of spontaneous autoimmune disease.
\newblock {\em The Journal of Immunology}, 167(9):5338--5347.

\bibitem[Esteller, 2008]{esteller2008epigenetics}
Esteller, M. (2008).
\newblock Epigenetics in cancer.
\newblock {\em New England Journal of Medicine}, 358(11):1148--1159.

\bibitem[Ferlay et~al., 2013]{ferlay2013cancer}
Ferlay, J., Steliarova-Foucher, E., Lortet-Tieulent, J., Rosso, S., Coebergh, J.-W.~W., Comber, H., Forman, D., and Bray, F. (2013).
\newblock Cancer incidence and mortality patterns in europe: estimates for 40 countries in 2012.
\newblock {\em European journal of cancer}, 49(6):1374--1403.

\bibitem[Gomez-Alonso et~al., 2021]{gomez2021dna}
Gomez-Alonso, M. d.~C., Kretschmer, A., Wilson, R., Pfeiffer, L., Karhunen, V., Sepp{\"a}l{\"a}, I., Zhang, W., Mittelstrass, K., Wahl, S., Matias-Garcia, P.~R., et~al. (2021).
\newblock Dna methylation and lipid metabolism: an ewas of 226 metabolic measures.
\newblock {\em Clinical epigenetics}, 13:1--19.

\bibitem[Habib et~al., 2022]{habib2022differential}
Habib, I., Anjum, F., Mohammad, T., Sulaimani, M.~N., Shafie, A., Almehmadi, M., Yadav, D.~K., Sohal, S.~S., and Hassan, M.~I. (2022).
\newblock Differential gene expression and network analysis in head and neck squamous cell carcinoma.
\newblock {\em Molecular and cellular biochemistry}, 477(5):1361--1370.

\bibitem[Hedl et~al., 2014]{hedl2014il18rap}
Hedl, M., Zheng, S., and Abraham, C. (2014).
\newblock The il18rap region disease polymorphism decreases il-18rap/il-18r1/il-1r1 expression and signaling through innate receptor--initiated pathways.
\newblock {\em The Journal of Immunology}, 192(12):5924--5932.

\bibitem[Herman and Baylin, 2003]{herman2003gene}
Herman, J.~G. and Baylin, S.~B. (2003).
\newblock Gene silencing in cancer in association with promoter hypermethylation.
\newblock {\em New England Journal of Medicine}, 349(21):2042--2054.

\bibitem[Herman et~al., 1996]{herman1996methylation}
Herman, J.~G., Graff, J.~R., My{\"o}h{\"a}nen, S., Nelkin, B.~D., and Baylin, S.~B. (1996).
\newblock Methylation-specific pcr: a novel pcr assay for methylation status of cpg islands.
\newblock {\em Proceedings of the national academy of sciences}, 93(18):9821--9826.

\bibitem[Herman et~al., 1994]{herman1994silencing}
Herman, J.~G., Latif, F., Weng, Y., Lerman, M.~I., Zbar, B., Liu, S., Samid, D., Duan, D., Gnarra, J.~R., and Linehan, W.~M. (1994).
\newblock Silencing of the vhl tumor-suppressor gene by dna methylation in renal carcinoma.
\newblock {\em Proceedings of the National Academy of Sciences}, 91(21):9700--9704.

\bibitem[Hirota et~al., 2012]{hirota2012genome}
Hirota, T., Takahashi, A., Kubo, M., Tsunoda, T., Tomita, K., Sakashita, M., Yamada, T., Fujieda, S., Tanaka, S., Doi, S., et~al. (2012).
\newblock Genome-wide association study identifies eight new susceptibility loci for atopic dermatitis in the japanese population.
\newblock {\em Nature genetics}, 44(11):1222--1226.

\bibitem[Hoffman et~al., 2001]{hoffman2001mutation}
Hoffman, H.~M., Mueller, J.~L., Broide, D.~H., Wanderer, A.~A., and Kolodner, R.~D. (2001).
\newblock Mutation of a new gene encoding a putative pyrin-like protein causes familial cold autoinflammatory syndrome and muckle--wells syndrome.
\newblock {\em Nature genetics}, 29(3):301--305.

\bibitem[Huang et~al., 2023]{huang2023higd2a}
Huang, K., Liu, Z., Xie, Z., Li, X., Zhang, H., Chen, Y., Wang, Y., Lin, Z., Li, C., Liu, H., et~al. (2023).
\newblock Higd2a silencing impairs hepatocellular carcinoma growth via inhibiting mitochondrial function and the mapk/erk pathway.
\newblock {\em Journal of Translational Medicine}, 21(1):253.

\bibitem[Huber et~al., 2012]{huber201222bp}
Huber, S., Gagliani, N., Zenewicz, L.~A., Huber, F.~J., Bosurgi, L., Hu, B., Hedl, M., Zhang, W., O’Connor, W., Murphy, A.~J., et~al. (2012).
\newblock Il-22bp is regulated by the inflammasome and modulates tumorigenesis in the intestine.
\newblock {\em Nature}, 491(7423):259--263.

\bibitem[Hunt et~al., 2008]{hunt2008newly}
Hunt, K.~A., Zhernakova, A., Turner, G., Heap, G.~A., Franke, L., Bruinenberg, M., Romanos, J., Dinesen, L.~C., Ryan, A.~W., Panesar, D., et~al. (2008).
\newblock Newly identified genetic risk variants for celiac disease related to the immune response.
\newblock {\em Nature genetics}, 40(4):395--402.

\bibitem[Iyer et~al., 2013]{iyer2013prevalence}
Iyer, G., Al-Ahmadie, H., Schultz, N., Hanrahan, A.~J., Ostrovnaya, I., Balar, A.~V., Kim, P.~H., Lin, O., Weinhold, N., Sander, C., et~al. (2013).
\newblock Prevalence and co-occurrence of actionable genomic alterations in high-grade bladder cancer.
\newblock {\em Journal of Clinical Oncology}, 31(25):3133--3140.

\bibitem[Jackson et~al., 2005]{jackson2005late}
Jackson, B., Tilli, C.~M., Hardman, M.~J., Avilion, A.~A., MacLeod, M.~C., Ashcroft, G.~S., and Byrne, C. (2005).
\newblock Late cornified envelope family in differentiating epithelia—response to calcium and ultraviolet irradiation.
\newblock {\em Journal of investigative dermatology}, 124(5):1062--1070.

\bibitem[Jen and Wang, 2016]{jen2016zinc}
Jen, J. and Wang, Y.-C. (2016).
\newblock Zinc finger proteins in cancer progression.
\newblock {\em Journal of biomedical science}, 23:1--9.

\bibitem[Jeon et~al., 2021]{jeon2021reliability}
Jeon, Y.-S., Shivakumar, M., Kim, D., Kim, C.-S., and Lee, J.-S. (2021).
\newblock Reliability of microarray analysis for studying periodontitis: low consistency in 2 periodontitis cohort data sets from different platforms and an integrative meta-analysis.
\newblock {\em Journal of periodontal \& implant science}, 51(1):18.

\bibitem[Jeronimo and Henrique, 2014]{jeronimo2014epigenetic}
Jeronimo, C. and Henrique, R. (2014).
\newblock Epigenetic biomarkers in urological tumors: A systematic review.
\newblock {\em Cancer letters}, 342(2):264--274.

\bibitem[Jiang et~al., 2012]{jiang2012role}
Jiang, M., Xu, S., Yue, W., Zhao, X., Zhang, L., Zhang, C., and Wang, Y. (2012).
\newblock The role of zfx in non-small cell lung cancer development.
\newblock {\em Oncology research}, 20(4):171--178.

\bibitem[Kamat et~al., 2016]{kamat2016bladder}
Kamat, A.~M., Hahn, N.~M., Efstathiou, J.~A., Lerner, S.~P., Malmstr{\"o}m, P.-U., Choi, W., Guo, C.~C., Lotan, Y., and Kassouf, W. (2016).
\newblock Bladder cancer.
\newblock {\em The Lancet}, 388(10061):2796--2810.

\bibitem[Kawakami et~al., 2000]{kawakami2000hypermethylated}
Kawakami, K., Brabender, J., Lord, R.~V., Groshen, S., Greenwald, B.~D., Krasna, M.~J., Yin, J., Fleisher, A.~S., Abraham, J.~M., Beer, D.~G., et~al. (2000).
\newblock Hypermethylated apc dna in plasma and prognosis of patients with esophageal adenocarcinoma.
\newblock {\em Journal of the National Cancer Institute}, 92(22):1805--1811.

\bibitem[Kawakami et~al., 2006]{kawakami2006identification}
Kawakami, K., Enokida, H., Tachiwada, T., Gotanda, T., Tsuneyoshi, K., Kubo, H., Nishiyama, K., Takiguchi, M., Nakagawa, M., and Seki, N. (2006).
\newblock Identification of differentially expressed genes in human bladder cancer through genome-wide gene expression profiling.
\newblock {\em Oncology reports}, 16(3):521--531.

\bibitem[Kawakami et~al., 2007]{kawakami2007increased}
Kawakami, K., Enokida, H., Tachiwada, T., Nishiyama, K., Seki, N., and Nakagawa, M. (2007).
\newblock Increased skp2 and cks1 gene expression contributes to the progression of human urothelial carcinoma.
\newblock {\em The Journal of urology}, 178(1):301--307.

\bibitem[Kawakami et~al., 1997]{kawakami199718}
Kawakami, K., Qureshi, M.~H., Zhang, T., Okamura, H., Kurimoto, M., and Saito, A. (1997).
\newblock Il-18 protects mice against pulmonary and disseminated infection with cryptococcus neoformans by inducing ifn-gamma production.
\newblock {\em Journal of immunology (Baltimore, Md.: 1950)}, 159(11):5528--5534.

\bibitem[Kikuchi et~al., 2009]{kikuchi2009trim24}
Kikuchi, M., Okumura, F., Tsukiyama, T., Watanabe, M., Miyajima, N., Tanaka, J., Imamura, M., and Hatakeyama, S. (2009).
\newblock Trim24 mediates ligand-dependent activation of androgen receptor and is repressed by a bromodomain-containing protein, brd7, in prostate cancer cells.
\newblock {\em Biochimica et Biophysica Acta (BBA)-Molecular Cell Research}, 1793(12):1828--1836.

\bibitem[Kim et~al., 2016]{kim2016role}
Kim, Y., Andr{\'e}s Salazar~Hern{\'a}ndez, M., Herrema, H., Delibasi, T., and Park, S.~W. (2016).
\newblock The role of brd 7 in embryo development and glucose metabolism.
\newblock {\em Journal of cellular and molecular medicine}, 20(8):1561--1570.

\bibitem[Lai et~al., 2014]{lai2014overexpression}
Lai, K.~P., Chen, J., He, M., Ching, A.~K., Lau, C., Lai, P.~B., To, K.-F., and Wong, N. (2014).
\newblock Overexpression of zfx confers self-renewal and chemoresistance properties in hepatocellular carcinoma.
\newblock {\em International journal of cancer}, 135(8):1790--1799.

\bibitem[Laird, 2005]{laird2005cancer}
Laird, P.~W. (2005).
\newblock Cancer epigenetics.
\newblock {\em Human molecular genetics}, 14(suppl\_1):R65--R76.

\bibitem[Lander, 2011]{lander2011initial}
Lander, E.~S. (2011).
\newblock Initial impact of the sequencing of the human genome.
\newblock {\em Nature}, 470(7333):187--197.

\bibitem[Langevin et~al., 2014]{langevin2014leukocyte}
Langevin, S.~M., Houseman, E.~A., Accomando, W.~P., Koestler, D.~C., Christensen, B.~C., Nelson, H.~H., Karagas, M.~R., Marsit, C.~J., Wiencke, J.~K., and Kelsey, K.~T. (2014).
\newblock Leukocyte-adjusted epigenome-wide association studies of blood from solid tumor patients.
\newblock {\em Epigenetics}, 9(6):884--895.

\bibitem[Lee et~al., 1997]{lee1997cg}
Lee, W.-H., Isaacs, W.~B., Bova, G.~S., and Nelson, W.~G. (1997).
\newblock {CG} island methylation changes near the {GSTP}1 gene in prostatic carcinoma cells detected using the polymerase chain reaction: a new prostate cancer biomarker.
\newblock {\em Cancer Epidemiology, Biomarkers \& Prevention}, 6(6):443--450.

\bibitem[Li et~al., 2003]{li20038}
Li, A., Dubey, S., Varney, M.~L., Dave, B.~J., and Singh, R.~K. (2003).
\newblock Il-8 directly enhanced endothelial cell survival, proliferation, and matrix metalloproteinases production and regulated angiogenesis.
\newblock {\em The Journal of Immunology}, 170(6):3369--3376.

\bibitem[Li et~al., 2013]{li2013zfx}
Li, K., Zhu, Z.-C., Liu, Y.-J., Liu, J.-W., Wang, H.-T., Xiong, Z.-Q., Shen, X., Hu, Z.-L., and Zheng, J. (2013).
\newblock Zfx knockdown inhibits growth and migration of non-small cell lung carcinoma cell line h1299.
\newblock {\em International journal of clinical and experimental pathology}, 6(11):2460.

\bibitem[Li et~al., 2023]{li2023brd7}
Li, M., Wei, Y., Liu, Y., Wei, J., Zhou, X., Duan, Y., Chen, S., Xue, C., Zhan, Y., Zheng, L., et~al. (2023).
\newblock Brd7 inhibits enhancer activity and expression of birc2 to suppress tumor growth and metastasis in nasopharyngeal carcinoma.
\newblock {\em Cell Death \& Disease}, 14(2):121.

\bibitem[Li et~al., 2022]{li2022structures}
Li, X., Han, M., Zhang, H., Liu, F., Pan, Y., Zhu, J., Liao, Z., Chen, X., and Zhang, B. (2022).
\newblock Structures and biological functions of zinc finger proteins and their roles in hepatocellular carcinoma.
\newblock {\em Biomarker research}, 10:1--13.

\bibitem[Li et~al., 2004]{li2004znf322}
Li, Y., Wang, Y., Zhang, C., Yuan, W., Wang, J., Zhu, C., Chen, L., Huang, W., Zeng, W., Wu, X., et~al. (2004).
\newblock Znf322, a novel human c2h2 kr{\"u}ppel-like zinc-finger protein, regulates transcriptional activation in mapk signaling pathways.
\newblock {\em Biochemical and biophysical research communications}, 325(4):1383--1392.

\bibitem[Li et~al., 2015]{li2015high}
Li, Y., Yan, X., Yan, L., Shan, Z., Liu, S., Chen, X., Zou, J., Zhang, W., and Jin, Z. (2015).
\newblock High expression of zinc-finger protein x-linked is associated with reduced e-cadherin expression and unfavorable prognosis in nasopharyngeal carcinoma.
\newblock {\em International journal of clinical and experimental pathology}, 8(4):3919.

\bibitem[Liang et~al., 2020]{liang2020mmp25}
Liang, Y., Guan, C., Li, K., Zheng, G., Wang, T., Zhang, S., and Liao, G. (2020).
\newblock Mmp25 regulates immune infiltration level and survival outcome in head and neck cancer patients.
\newblock {\em Frontiers in Oncology}, 10:1088.

\bibitem[Liep et~al., 2012]{liep2012feedback}
Liep, J., Rabien, A., and Jung, K. (2012).
\newblock Feedback networks between micrornas and epigenetic modifications in urological tumors.
\newblock {\em Epigenetics}, 7(4):315--325.

\bibitem[Liu et~al., 2012]{liu2012identification}
Liu, H., Irwanto, A., Tian, H., Yu, Y., Yu, G., Low, H., Chu, T., Li, Y., Shi, B., Chen, M., et~al. (2012).
\newblock Identification of il18rap/il18r1 and il12b as leprosy risk genes demonstrates shared pathogenesis between inflammation and infectious diseases.
\newblock {\em The American Journal of Human Genetics}, 91(5):935--941.

\bibitem[Liu et~al., 2015a]{liu2015baicalein}
Liu, T.-Y., Gong, W., Tan, Z.-J., Lu, W., Wu, X.-S., Weng, H., Ding, Q., Shu, Y.-J., Bao, R.-F., Cao, Y., et~al. (2015a).
\newblock Baicalein inhibits progression of gallbladder cancer cells by downregulating zfx.
\newblock {\em PloS one}, 10(1):e0114851.

\bibitem[Liu et~al., 2016]{liu2016mir}
Liu, Y., Zhao, R., Wang, H., Luo, Y., Wang, X., Niu, W., Zhou, Y., Wen, Q., Fan, S., Li, X., et~al. (2016).
\newblock mir-141 is involved in brd7-mediated cell proliferation and tumor formation through suppression of the pten/akt pathway in nasopharyngeal carcinoma.
\newblock {\em Cell death \& disease}, 7(3):e2156--e2156.

\bibitem[Liu et~al., 2018]{liu2018brd7}
Liu, Y., Zhao, R., Wei, Y., Li, M., Wang, H., Niu, W., Zhou, Y., Qiu, Y., Fan, S., Zhan, Y., et~al. (2018).
\newblock Brd7 expression and c-myc activation forms a double-negative feedback loop that controls the cell proliferation and tumor growth of nasopharyngeal carcinoma by targeting oncogenic mir-141.
\newblock {\em Journal of Experimental \& Clinical Cancer Research}, 37:1--14.

\bibitem[Liu et~al., 2015b]{liu2015dach1}
Liu, Y., Zhou, R., Yuan, X., Han, N., Zhou, S., Xu, H., Guo, M., Yu, S., Zhang, C., Yin, T., et~al. (2015b).
\newblock Dach1 is a novel predictive and prognostic biomarker in hepatocellular carcinoma as a negative regulator of wnt/$\beta$-catenin signaling.
\newblock {\em Oncotarget}, 6(11):8621.

\bibitem[Martinez et~al., 2019]{martinez2019epigenetics}
Martinez, V.~G., Munera-Maravilla, E., Bernardini, A., Rubio, C., Suarez-Cabrera, C., Segovia, C., Lodewijk, I., Due{\~n}as, M., Mart{\'\i}nez-Fern{\'a}ndez, M., and Paramio, J.~M. (2019).
\newblock Epigenetics of bladder cancer: where biomarkers and therapeutic targets meet.
\newblock {\em Frontiers in Genetics}, 10:1125.

\bibitem[Martinon et~al., 2002]{martinon2002inflammasome}
Martinon, F., Burns, K., and Tschopp, J. (2002).
\newblock The inflammasome: a molecular platform triggering activation of inflammatory caspases and processing of proil-$\beta$.
\newblock {\em Molecular cell}, 10(2):417--426.

\bibitem[Martinon et~al., 2007]{martinon2007nalp}
Martinon, F., Gaide, O., P{\'e}trilli, V., Mayor, A., and Tschopp, J. (2007).
\newblock Nalp inflammasomes: a central role in innate immunity.
\newblock In {\em Seminars in immunopathology}, volume~29, pages 213--229. Springer.

\bibitem[Martinon and Tschopp, 2005]{martinon2005nlrs}
Martinon, F. and Tschopp, J. (2005).
\newblock Nlrs join tlrs as innate sensors of pathogens.
\newblock {\em Trends in immunology}, 26(8):447--454.

\bibitem[Mitani et~al., 2000]{mitani2000nonredundant}
Mitani, K., Yamagata, T., Iida, C., Oda, H., Maki, K., Ichikawa, M., Asai, T., Honda, H., Kurokawa, M., and Hirai, H. (2000).
\newblock Nonredundant roles of the elongation factor men in postimplantation development.
\newblock {\em Biochemical and biophysical research communications}, 279(2):563--567.

\bibitem[Mitra et~al., 2006]{mitra2006molecular}
Mitra, A.~P., Datar, R.~H., and Cote, R.~J. (2006).
\newblock Molecular pathways in invasive bladder cancer: new insights into mechanisms, progression, and target identification.
\newblock {\em Journal of Clinical Oncology}, 24(35):5552--5564.

\bibitem[Nagarajan et~al., 2009]{nagarajan2009role}
Nagarajan, P., Onami, T.~M., Rajagopalan, S., Kania, S., Donnell, R., and Venkatachalam, S. (2009).
\newblock Role of chromodomain helicase dna-binding protein 2 in dna damage response signaling and tumorigenesis.
\newblock {\em Oncogene}, 28(8):1053--1062.

\bibitem[Nathan, 2006]{nathan2006neutrophils}
Nathan, C. (2006).
\newblock Neutrophils and immunity: challenges and opportunities.
\newblock {\em Nature reviews immunology}, 6(3):173--182.

\bibitem[Niu et~al., 2020]{niu2020brd7}
Niu, W., Luo, Y., Zhou, Y., Li, M., Wu, C., Duan, Y., Wang, H., Fan, S., Li, Z., Xiong, W., et~al. (2020).
\newblock Brd7 suppresses invasion and metastasis in breast cancer by negatively regulating yb1-induced epithelial-mesenchymal transition.
\newblock {\em Journal of Experimental \& Clinical Cancer Research}, 39:1--19.

\bibitem[Park and Lee, 2020]{park2020emerging}
Park, S.~W. and Lee, J.~M. (2020).
\newblock Emerging roles of brd7 in pathophysiology.
\newblock {\em International Journal of Molecular Sciences}, 21(19):7127.

\bibitem[Park et~al., 2014]{park2014tumor}
Park, Y.-A., Lee, J.-W., Kim, H.-S., Lee, Y.-Y., Kim, T.-J., Choi, C.~H., Choi, J.-J., Jeon, H.-K., Cho, Y.~J., Ryu, J.~Y., et~al. (2014).
\newblock Tumor suppressive effects of bromodomain-containing protein 7 (brd7) in epithelial ovarian carcinoma.
\newblock {\em Clinical Cancer Research}, 20(3):565--575.

\bibitem[Peng et~al., 2007]{peng2007brd7}
Peng, C., Liu, H.~Y., Zhou, M., Zhang, L.~M., Li, X.~L., Shen, S.~R., and Li, G.~Y. (2007).
\newblock Brd7 suppresses the growth of nasopharyngeal carcinoma cells (hne1) through negatively regulating $\beta$-catenin and erk pathways.
\newblock {\em Molecular and cellular biochemistry}, 303:141--149.

\bibitem[Peng et~al., 2002]{peng2002analysis}
Peng, C., Zhou, M., Zhang, Q., Tang, K., and Li, G. (2002).
\newblock Analysis of bromodomain of brd-7 gene and its prokaryotic expression.
\newblock {\em Ai Zheng= Aizheng= Chinese Journal of Cancer}, 21(11):1167--1172.

\bibitem[Penkert et~al., 2012]{penkert2012no}
Penkert, J., Schlegelberger, B., Steinemann, D., and Gadzicki, D. (2012).
\newblock No evidence for breast cancer susceptibility associated with variants of brd7, a component of p53 and brca1 pathways.
\newblock {\em Familial cancer}, 11:601--606.

\bibitem[Pidsley et~al., 2016]{pidsley2016critical}
Pidsley, R., Zotenko, E., Peters, T.~J., Lawrence, M.~G., Risbridger, G.~P., Molloy, P., Van~Djik, S., Muhlhausler, B., Stirzaker, C., and Clark, S.~J. (2016).
\newblock Critical evaluation of the {I}llumina {M}ethylationepic {B}ead{C}hip microarray for whole-genome {DNA} methylation profiling.
\newblock {\em Genome {B}iology}, 17(1):1--17.

\bibitem[Popov et~al., 2010]{popov2010dachshund}
Popov, V.~M., Wu, K., Zhou, J., Powell, M.~J., Mardon, G., Wang, C., and Pestell, R.~G. (2010).
\newblock The dachshund gene in development and hormone-responsive tumorigenesis.
\newblock {\em Trends in Endocrinology \& Metabolism}, 21(1):41--49.

\bibitem[Porten, 2018]{porten2018epigenetic}
Porten, S.~P. (2018).
\newblock Epigenetic alterations in bladder cancer.
\newblock {\em Current Urology Reports}, 19:1--8.

\bibitem[Powe et~al., 2014]{powe2014dach1}
Powe, D.~G., Dhondalay, G. K.~R., Lemetre, C., Allen, T., Habashy, H.~O., Ellis, I.~O., Rees, R., and Ball, G.~R. (2014).
\newblock Dach1: its role as a classifier of long term good prognosis in luminal breast cancer.
\newblock {\em PloS one}, 9(1):e84428.

\bibitem[Rakyan et~al., 2011]{rakyan2011epigenome}
Rakyan, V.~K., Down, T.~A., Balding, D.~J., and Beck, S. (2011).
\newblock Epigenome-wide association studies for common human diseases.
\newblock {\em Nature Reviews Genetics}, 12(8):529--541.

\bibitem[Ransohoff, 2003]{ransohoff2003developing}
Ransohoff, D.~F. (2003).
\newblock Developing molecular biomarkers for cancer.
\newblock {\em Science}, 299(5613):1679--1680.

\bibitem[Ricciardiello et~al., 2003]{ricciardiello2003frequent}
Ricciardiello, L., Goel, A., Mantovani, V., Fiorini, T., Fossi, S., Chang, D.~K., Lunedei, V., Pozzato, P., Zagari, R.~M., De~Luca, L., et~al. (2003).
\newblock Frequent loss of hmlh1 by promoter hypermethylation leads to microsatellite instability in adenomatous polyps of patients with a single first-degree member affected by colon cancer.
\newblock {\em Cancer research}, 63(4):787--792.

\bibitem[R{\"o}nn and Ling, 2015]{ronn2015dna}
R{\"o}nn, T. and Ling, C. (2015).
\newblock {DNA} methylation as a diagnostic and therapeutic target in the battle against {T}ype 2 {D}iabetes.
\newblock {\em Epigenomics}, 7(3):451--460.

\bibitem[Rosen et~al., 2018]{rosen2018epigenetics}
Rosen, E.~D., Kaestner, K.~H., Natarajan, R., Patti, M.-E., Sallari, R., Sander, M., and Susztak, K. (2018).
\newblock Epigenetics and epigenomics: implications for {D}iabetes and obesity.
\newblock {\em {D}iabetes}, 67(10):1923--1931.

\bibitem[Salazar et~al., 2020]{salazar2020biosystem}
Salazar, C., Ya{\~n}ez, O., Elorza, A.~A., Cortes, N., Garc{\'\i}a-Beltr{\'a}n, O., Tiznado, W., and Ruiz, L.~M. (2020).
\newblock Biosystem analysis of the hypoxia inducible domain family member 2a: Implications in cancer biology.
\newblock {\em Genes}, 11(2):206.

\bibitem[Santourlidis et~al., 1999]{santourlidis1999high}
Santourlidis, S., Florl, A., Ackermann, R., Wirtz, H.-C., and Schulz, W.~A. (1999).
\newblock High frequency of alterations in dna methylation in adenocarcinoma of the prostate.
\newblock {\em The Prostate}, 39(3):166--174.

\bibitem[Sharma et~al., 2010]{sharma2010epigenetics}
Sharma, S., Kelly, T.~K., and Jones, P.~A. (2010).
\newblock Epigenetics in cancer.
\newblock {\em Carcinogenesis}, 31(1):27--36.

\bibitem[Siva et~al., 2008]{siva2008immune}
Siva, A., Xin, H., Qin, F., Oltean, D., Bowdish, K., and Kretz-Rommel, A. (2008).
\newblock Immune modulation by melanoma and ovarian tumor cells through expression of the immunosuppressive molecule cd200.
\newblock {\em Cancer Immunology, Immunotherapy}, 57:987--996.

\bibitem[Sjoblom et~al., 2006]{sjoblom2006consensus}
Sjoblom, T., Jones, S., Wood, L.~D., Parsons, D.~W., Lin, J., Barber, T.~D., Mandelker, D., Leary, R.~J., Ptak, J., Silliman, N., et~al. (2006).
\newblock The consensus coding sequences of human breast and colorectal cancers.
\newblock {\em science}, 314(5797):268--274.

\bibitem[Smyth et~al., 2008]{smyth2008shared}
Smyth, D.~J., Plagnol, V., Walker, N.~M., Cooper, J.~D., Downes, K., Yang, J.~H., Howson, J.~M., Stevens, H., McManus, R., Wijmenga, C., et~al. (2008).
\newblock Shared and distinct genetic variants in type 1 diabetes and celiac disease.
\newblock {\em New England Journal of Medicine}, 359(26):2767--2777.

\bibitem[Society, 2024]{acs2024bladder}
Society, A.~C. (2024).
\newblock Key statistics for bladder cancer.
\newblock Accessed on August 18, 2024.

\bibitem[Sondka et~al., 2018]{sondka2018cosmic}
Sondka, Z., Bamford, S., Cole, C.~G., Ward, S.~A., Dunham, I., and Forbes, S.~A. (2018).
\newblock The cosmic cancer gene census: describing genetic dysfunction across all human cancers.
\newblock {\em Nature Reviews Cancer}, 18(11):696--705.

\bibitem[Stadler et~al., 2010]{stadler2010genome}
Stadler, Z.~K., Vijai, J., Thom, P., Kirchhoff, T., Hansen, N.~A., Kauff, N.~D., Robson, M., and Offit, K. (2010).
\newblock Genome-wide association studies of cancer predisposition.
\newblock {\em Hematology/Oncology Clinics}, 24(5):973--996.

\bibitem[Stanley et~al., 2013]{stanley2013chd}
Stanley, F.~K., Moore, S., and Goodarzi, A.~A. (2013).
\newblock Chd chromatin remodelling enzymes and the dna damage response.
\newblock {\em Mutation Research/Fundamental and Molecular Mechanisms of Mutagenesis}, 750(1-2):31--44.

\bibitem[Stevens et~al., 2017]{stevens2017comparison}
Stevens, J.~R., Al~Masud, A., and Suyundikov, A. (2017).
\newblock A comparison of multiple testing adjustment methods with block-correlation positively-dependent tests.
\newblock {\em PloS One}, 12(4):e0176124.

\bibitem[Susan et~al., 1994]{susan1994high}
Susan, J.~C., Harrison, J., Paul, C.~L., and Frommer, M. (1994).
\newblock High sensitivity mapping of methylated cytosines.
\newblock {\em Nucleic acids research}, 22(15):2990--2997.

\bibitem[Sweta and Sharma, 2021]{sweta2021functional}
Sweta, K. and Sharma, N. (2021).
\newblock Functional interaction between ell transcription elongation factor and epe1 reveals the role of epe1 in the regulation of transcription outside heterochromatin.
\newblock {\em Molecular Microbiology}, 116(1):80--96.

\bibitem[Thirman et~al., 1994]{thirman1994cloning}
Thirman, M.~J., Levitan, D.~A., Kobayashi, H., Simon, M.~C., and Rowley, J.~D. (1994).
\newblock Cloning of ell, a gene that fuses to mll in at (11; 19)(q23; p13. 1) in acute myeloid leukemia.
\newblock {\em Proceedings of the National Academy of Sciences}, 91(25):12110--12114.

\bibitem[Tschopp et~al., 2003]{tschopp2003nalps}
Tschopp, J., Martinon, F., and Burns, K. (2003).
\newblock Nalps: a novel protein family involved in inflammation.
\newblock {\em Nature reviews Molecular cell biology}, 4(2):95--104.

\bibitem[Wang et~al., 2016]{wang2016long}
Wang, T., Ma, S., Qi, X., Tang, X., Cui, D., Wang, Z., Chi, J., Li, P., and Zhai, B. (2016).
\newblock Long noncoding rna znfx1-as1 suppresses growth of hepatocellular carcinoma cells by regulating the methylation of mir-9.
\newblock {\em OncoTargets and therapy}, pages 5005--5014.

\bibitem[Waugh and Wilson, 2008]{waugh2008interleukin}
Waugh, D.~J. and Wilson, C. (2008).
\newblock The interleukin-8 pathway in cancer.
\newblock {\em Clinical cancer research}, 14(21):6735--6741.

\bibitem[Weng et~al., 2015]{weng2015zinc}
Weng, H., Wang, X., Li, M., Wu, X., Wang, Z., Wu, W., Zhang, Z., Zhang, Y., Zhao, S., Liu, S., et~al. (2015).
\newblock Zinc finger x-chromosomal protein (zfx) is a significant prognostic indicator and promotes cellular malignant potential in gallbladder cancer.
\newblock {\em Cancer biology \& therapy}, 16(10):1462--1470.

\bibitem[Wolffe and Matzke, 1999]{wolffe1999epigenetics}
Wolffe, A.~P. and Matzke, M.~A. (1999).
\newblock Epigenetics: regulation through repression.
\newblock {\em science}, 286(5439):481--486.

\bibitem[Wu et~al., 2014]{wu2014cell}
Wu, K., Chen, K., Wang, C., Jiao, X., Wang, L., Zhou, J., Wang, J., Li, Z., Addya, S., Sorensen, P.~H., et~al. (2014).
\newblock Cell fate factor dach1 represses yb-1--mediated oncogenic transcription and translation.
\newblock {\em Cancer research}, 74(3):829--839.

\bibitem[Wu et~al., 2011]{wu2011cell}
Wu, K., Jiao, X., Li, Z., Katiyar, S., Casimiro, M.~C., Yang, W., Zhang, Q., Willmarth, N.~E., Chepelev, I., Crosariol, M., et~al. (2011).
\newblock Cell fate determination factor dachshund reprograms breast cancer stem cell function.
\newblock {\em Journal of Biological Chemistry}, 286(3):2132--2142.

\bibitem[Wu et~al., 2006]{wu2006dach1}
Wu, K., Li, A., Rao, M., Liu, M., Dailey, V., Yang, Y., Di~Vizio, D., Wang, C., Lisanti, M.~P., Sauter, G., et~al. (2006).
\newblock Dach1 is a cell fate determination factor that inhibits cyclin d1 and breast tumor growth.
\newblock {\em Molecular and Cellular Biology}, 26(19):7116--7129.

\bibitem[Wu et~al., 2013a]{wu2013knockdown}
Wu, S., Lao, X.-Y., Sun, T.-T., Ren, L.-L., Kong, X., Wang, J.-L., Wang, Y.-C., Du, W., Yu, Y.-N., Weng, Y.-R., et~al. (2013a).
\newblock Knockdown of zfx inhibits gastric cancer cell growth in vitro and in vivo via downregulating the erk-mapk pathway.
\newblock {\em Cancer letters}, 337(2):293--300.

\bibitem[Wu et~al., 2015]{wu2015six1}
Wu, W., Ren, Z., Li, P., Yu, D., Chen, J., Huang, R., and Liu, H. (2015).
\newblock Six1: a critical transcription factor in tumorigenesis.
\newblock {\em International journal of cancer}, 136(6):1245--1253.

\bibitem[Wu et~al., 2013b]{wu2013prognostic}
Wu, W.-J., Hu, K.-S., Chen, D.-L., Zeng, Z.-L., Luo, H.-Y., Wang, F., Wang, D.-S., Wang, Z.-Q., He, F., and Xu, R.-H. (2013b).
\newblock Prognostic relevance of brd 7 expression in colorectal carcinoma.
\newblock {\em European journal of clinical investigation}, 43(2):131--140.

\bibitem[Yan et~al., 2022]{yan2022transcriptomic}
Yan, B., Ren, F., Shang, W., and Gong, X. (2022).
\newblock Transcriptomic analysis reveals genetic cross-talk between periodontitis and hypothyroidism.
\newblock {\em Disease Markers}, 2022(1):5736394.

\bibitem[Yan et~al., 2013]{yan2013epigenetic}
Yan, W., Wu, K., Herman, J.~G., Brock, M.~V., Fuks, F., Yang, L., Zhu, H., Li, Y., Yang, Y., and Guo, M. (2013).
\newblock Epigenetic regulation of dach1, a novel wnt signaling component in colorectal cancer.
\newblock {\em Epigenetics}, 8(12):1373--1383.

\bibitem[Yang et~al., 2014]{yang2014shrna}
Yang, H., Lu, Y., Zheng, Y., Yu, X., Xia, X., He, X., Feng, W., Xing, L., and Ling, Z. (2014).
\newblock shrna-mediated silencing of zfx attenuated the proliferation of breast cancer cells.
\newblock {\em Cancer chemotherapy and pharmacology}, 73:569--576.

\bibitem[Yang et~al., 2010]{yang2010targeting}
Yang, X., Lay, F., Han, H., and Jones, P.~A. (2010).
\newblock Targeting {DNA} methylation for epigenetic therapy.
\newblock {\em Trends in Pharmacological Sciences}, 31(11):536--546.

\bibitem[Yoshimura et~al., 2020]{yoshimura2020cd200}
Yoshimura, K., Suzuki, Y., Inoue, Y., Tsuchiya, K., Karayama, M., Iwashita, Y., Kahyo, T., Kawase, A., Tanahashi, M., Ogawa, H., et~al. (2020).
\newblock Cd200 and cd200r1 are differentially expressed and have differential prognostic roles in non-small cell lung cancer.
\newblock {\em Oncoimmunology}, 9(1):1746554.

\bibitem[Yu et~al., 2016]{yu2016brd7}
Yu, X., Li, Z., and Shen, J. (2016).
\newblock Brd7: a novel tumor suppressor gene in different cancers.
\newblock {\em American journal of translational research}, 8(2):742.

\bibitem[Yu et~al., 2000]{yu2000analysis}
Yu, Y., Zhang, B.-C., Xie, Y., Cao, L., Zhou, M., Zhan, F.-H., and Li, G.-Y. (2000).
\newblock Analysis and molecular cloning of differentially expressing genes in nasopharyngeal carcinoma.
\newblock {\em Sheng wu hua xue yu Sheng wu wu li xue bao Acta Biochimica et Biophysica Sinica}, 32(4):327--332.

\bibitem[Zaki et~al., 2010]{zaki2010nlrp3}
Zaki, M.~H., Boyd, K.~L., Vogel, P., Kastan, M.~B., Lamkanfi, M., and Kanneganti, T.-D. (2010).
\newblock The nlrp3 inflammasome protects against loss of epithelial integrity and mortality during experimental colitis.
\newblock {\em Immunity}, 32(3):379--391.

\bibitem[Zhang et~al., 2003]{zhang2003expression}
Zhang, X.-M., Sheng, S.-R., Wang, X.-Y., Wang, J.-R., and Li, J. (2003).
\newblock Expression of tumor related genes ngx6, nag-7, brd7 in gastric and colorectal cancer.
\newblock {\em World Journal of Gastroenterology: WJG}, 9(8):1729.

\bibitem[Zhang et~al., 2021]{zhang2021identification}
Zhang, Y.-Y., Mao, M.-H., and Han, Z.-X. (2021).
\newblock Identification of a gene prognostic signature for oral squamous cell carcinoma by rna sequencing and bioinformatics.
\newblock {\em BioMed Research International}, 2021(1):6657767.

\bibitem[Zhernakova et~al., 2008]{zhernakova2008genetic}
Zhernakova, A., Festen, E.~M., Franke, L., Trynka, G., van Diemen, C.~C., Monsuur, A.~J., Bevova, M., Nijmeijer, R.~M., van‘t Slot, R., Heijmans, R., et~al. (2008).
\newblock Genetic analysis of innate immunity in crohn's disease and ulcerative colitis identifies two susceptibility loci harboring card9 and il18rap.
\newblock {\em The American Journal of Human Genetics}, 82(5):1202--1210.

\bibitem[Zhou et~al., 2004]{zhou2004brd7}
Zhou, J., Ma, J., Zhang, B.-C., Li, X.-L., Shen, S.-R., Zhu, S.-G., Xiong, W., Liu, H.-Y., Huang, H., Zhou, M., et~al. (2004).
\newblock Brd7, a novel bromodomain gene, inhibits g1--s progression by transcriptionally regulating some important molecules involved in ras/mek/erk and rb/e2f pathways.
\newblock {\em Journal of cellular physiology}, 200(1):89--98.

\bibitem[Zhou et~al., 2006]{zhou2006identification}
Zhou, M., Liu, H., Xu, X., Zhou, H., Li, X., Peng, C., Shen, S., Xiong, W., Ma, J., Zeng, Z., et~al. (2006).
\newblock Identification of nuclear localization signal that governs nuclear import of brd7 and its essential roles in inhibiting cell cycle progression.
\newblock {\em Journal of Cellular Biochemistry}, 98(4):920--930.

\bibitem[Zhu et~al., 2013]{zhu2013epigenetic}
Zhu, H., Wu, K., Yan, W., Hu, L., Yuan, J., Dong, Y., Li, Y., Jing, K., Yang, Y., and Guo, M. (2013).
\newblock Epigenetic silencing of dach1 induces loss of transforming growth factor-$\beta$1 antiproliferative response in human hepatocellular carcinoma.
\newblock {\em Hepatology}, 58(6):2012--2022.

\end{thebibliography}

\end{document}